\documentclass[12pt]{article}
\usepackage{amssymb}
\usepackage{amsmath}
\usepackage{amsthm}
\usepackage{cancel}
\usepackage{slashed}
\usepackage{fullpage}
\usepackage{color}
\usepackage{braket}
\usepackage{graphicx}
\usepackage[small]{subfigure}
\usepackage{multirow}
\usepackage{verbatim}
\usepackage{cite}
\usepackage{bigints}
\usepackage{simplewick}
\usepackage{authblk}
\usepackage{hyperref}
\usepackage{empheq}
\usepackage{tikz-feynman}
\usepackage{tikz}

\hypersetup{
    linktocpage,
     colorlinks,
     citecolor=darkgreen,
     linkcolor= darkgreen,
     urlcolor=darkgreen
}

\definecolor{darkred}{rgb}{0.5,0.0,0.0}
\definecolor{darkblue}{rgb}{0.0,0.0,0.9}
\definecolor{darkerblue}{rgb}{0.0,0.0,0.5}
\definecolor{darkgreen}{rgb}{0.0,0.5,0.0}
\definecolor{black}{rgb}{0.0,0.0,0.0}
\definecolor{brown}{rgb}{0.6,0.4,0.2}

\numberwithin{equation}{section}

\title{Light dark matter from dark sector decay}

\author{Yu Cheng$^{1,2}$\thanks{chengyu@mail.ecust.edu.cn}}
\author{Wei Liao$^1$\thanks{liaow@ecust.edu.cn}}

\affil{\emph{\vskip 0.2cm
		${}^{1}$Institute of Modern Physics,  School of Sciences,\\
		East China University of Science and Technology, 130 Meilong Road, Shanghai 200237, P. R. China}}
\affil{\emph{\vskip -0.5cm ${}^{2}$Tsung-Dao Lee Institute, and School of Physics and Astronomy, \\
Shanghai Jiao Tong University, Shanghai 200240, P. R. China}}

\begin{document} 
\maketitle

\begin{abstract}
We study the possibility that light dark matter can be produced with right relic density by the decay of other dark sector
particles. We study this possibility in a model with right-handed neutrinos, a dark sector singlet scalar and a dark sector
singlet fermion. The decay of the heavier dark sector singlet scalar gives rise to the lighter dark sector singlet fermion 
which serves as the dark matter candidate. We show that the right dark matter relic density can be produced by the decay 
of  the dark sector singlet scalar. We find that the mass of the dark sector singlet fermion  can be GeV scale or MeV scale 
and the interaction of the dark sector singlet fermion is very weak. The dark sector singlet scalar can have a mass of GeV-TeV scale 
 and can have weak scale interactions with the Standard Model(SM) particles. So, in this scenario, dark matter  has an 
interaction with the SM much weaker than the weak interaction,  but the dark sector still has weak scale interaction with the SM.

\end{abstract}

\section{Introduction}
The evidences of Dark Matter(DM) in cosmological and astronomical observations convincingly suggest that there are physics beyond 
the Standard Model(SM) of particle physics. 
Candidate of  DM particle should be stable or have life-time much longer than the age of the Universe. 
It should not  carry electromagnetic charge so that it does not have long range interaction with ordinary matter.
According to cosmological observation,  DM should account for around 23\% of the energy budget of the Universe.
Weakly Interacting Massive Particles(WIMPs) have masses of GeV scale  and a strength of interaction of the weak interaction,
and have been considered as a good candidate of DM.
A surprising feature of WIMP DM is that the annihilation of WIMP particles 
when freezing out in the thermal early universe can naturally give rise to a relic energy density
consistent with the relic DM energy density discovered in cosmological observation.  This is the so-called WIMP miracle.

Many experiments have been done to search for the direct or indirect signals of DM particles
~\cite{Aprile:2018dbl,cui2017dark,Wang:2020coa,Akerib:2016vxi,Campos:2017odj,Aartsen:2014hva,Khachatryan:2014rra,ATLAS:2012ky} .
Among them, a lot of attentions  have been paid to the  detection of WIMP DM. 
So far, no signals of DM particles have been found in these experiments.
In particular, the direct detection experiments~\cite{Aprile:2018dbl,cui2017dark,Wang:2020coa} have pushed  
WIMP-nucleus interaction cross section to be less than around $10^{-46}$ cm$^2$ . 
These direct detection measurements have raised serious concern that WIMP should have interactions much weaker than we thought before
and over-production of WIMP relic density in thermal production mechanism seems hard to avoid.

One way to overcome this difficulty is to consider other DM production mechanisms and to consider DM candidates not in GeV mass range.
One possible production mechanism is the freeze-in mechanism~\cite{hall2010freeze}. 
This  mechanism assumes that the coupling of the dark sector and SM particles is so weak
that DM never get into chemical equilibrium with the SM thermal bath in the early universe.
Consequently,  the DM particles can be produced by the annihilation of other particles in the early universe,
but the production is so slow that only a small number density can be accumulated.
Another possibility is that DM particle may appear the decay product of  other particles which, on the other hand, can have a relic density
obtained with a thermal production mechanism. As long as the mass of the former is much smaller than the mass of the later, 
the problem of over-production in thermal production can be  significantly eased.
In this scenario, the DM should also have very weak interaction with the SM particles so that thermal production mechanism does not apply.

In this paper,  we consider this possibility that DM particle is the decay product of other heavier particle in the dark sector.
The DM particles never get into thermal and chemical equilibrium in the early universe,
but the heavier particles are in the thermal equilibrium with the SM particles  and can obtain a relic density when they freeze out. 
Then, heavier particles decay to the DM particle which can be much lighter than the former.
Since the DM particle has very weak interaction with SM particles, the freeze-in mechanism can also contribute to the DM relic density.
The advantage of this scenario is that the mass region of the dark sector particles can be 
widely expanded and the constraints on the parameter space is much weaker.

As an example, we consider a DM model in which the right-handed neutrino(RHN) works as a 
portal\cite{Macias:2015cna,blennow2019neutrino,escudero2017sterile,macias2016realistic} to the DM. 
Previous works on this type of models have focused on the possibility that the DM relic density 
is achieved by the thermal freeze-out mechanism\cite{escudero2017sterile,macias2016realistic,Bandyopadhyay:2018qcv} or by the freeze-in mechanism\cite{Chianese:2018dsz,Bandyopadhyay:2020qpn,Chianese:2019epo,Chianese:2020khl}.
We consider the possibility that the DM relic density is given mainly by the decay of other heavier particles in the dark sector 
which obtain a relic density by the thermal freeze-out mechanism. 

The rest of the paper is organized as follows.
In Sec.~\ref{sec:model} we introduce the model. 
In Sec.~\ref{sec:freezein} we discuss in detail the production of DM relic density in this model.
In Sec.\ref{sec:numerical} we provide some numerical results for the DM relic abundance 
and present constraints on  the parameter space. We conclude in the last section.
 
\section{The model of DM}
\label{sec:model}
We consider an extension of  the Type-I seesaw model where the  right-handed neutrinos $N_{R}$ are coupled to a dark sector 
consisting of a real scalar $\phi$ and a Dirac fermion $\chi$. 
We impose a $Z_2$ symmetry in the Lagrangian with the dark sector particles $\chi$ and $\phi$ odd and SM fermions even under the $Z_2$ operation. 
This guarantees that the lightest of the dark sector particles to be stable, and makes it a DM candidate. 
We can write the full Lagrangian as the sum of four parts
\begin{equation}
\mathcal{L}=\mathcal{L}_{\mathrm{SM}}+\mathcal{L}_{\text {Seesaw }}+\mathcal{L}_{\mathrm{DS}}+\mathcal{L}_{\text {int}}
\end{equation}
where the first term is the SM Lagrangian and the other three terms are as follows.
\begin{eqnarray}
\mathcal{L}_{\text {Seesaw }}&&=-Y_{\alpha \beta} \overline{L^{\alpha}} \tilde{H} N_{R \beta}-\frac{1}{2} M_{N} \overline{N_{R}^{c}} N_{R}+\text { h.c. } \\
\mathcal{L}_{\text {DS }}&&=\bar{\chi}\left(i \not\partial-m_{\chi}\right) \chi + \frac{1}{2} \partial_{\mu} \phi \partial^{\mu} \phi-\frac{1}{2}\mu_{\phi}^{2} \phi^{2} -\frac{1}{4}\lambda_{\phi} {\phi}^4 
-\lambda_{H \phi}\left(H^{\dagger} H\right)\phi^{2} .\\
\mathcal{L}_{\text {int }} &&=-\left( y_{\text {DS }} \phi \bar{\chi} N_{R}+h.c \right).
\end{eqnarray}
In these equations, $L^\alpha$ is the left-handed lepton doublets, $H$  the SM Higgs doublet with $\tilde{H}=i \tau_{2} H^{*}$,
$m_{\chi}$ the mass of $\chi$, $M_N$ the mass of the right-handed neutrinos. 
There are three generations of right-handed neutrinos and we consider the simplest case that 
the masses of the three generations of right-handed neutrinos are the same.
After the electroweak symmetry breaking, the scalar $\phi$ gets a further contribution to its mass and
we have 
\begin{equation}
m^2_{\phi} = {\mu}^2_{\phi}+\lambda_{H \phi} v_{ew}^2,
\end{equation} 
 where  $v_{ew}/\sqrt{2}$ is the vacuum expectation of Higgs doublet $H$ after the electroweak symmetry breaking.

In this model, the scalar $\phi$ in the dark sector couples to the SM sector directly with the coupling constant $\lambda_{H\phi}$.
The fermion $\chi$ in the dark sector couples directly to the right-handed neutrinos and the scalar $\phi$ with a coupling constant $y_{DS}$,
but does not couple to the SM sector directly.

\section{The production of the DM relic density}
\label{sec:freezein}
In this model, both the scalar and the fermion can be the DM candidate. 
In this paper, we focus on the case with fermion DM which is produced mainly through heavy dark scalar decay.

We assume $\phi$ is heavier than the fermion $\chi$ and the right-handed neutrinos $N$, and 
$\phi$ can decay to $\chi$ plus $N$.
We assume that  the coupling $\lambda_{H\phi}$ is not small so that the scalar $\phi$ is in the thermal equilibrium with the SM particles at high temperature. 
As temperature drops down to the mass scale of $\phi$,  the heavier $\phi$  freezes out and decouples from the thermal bath.
Then,  $\phi$ after freeze-out decays  into fermion $\chi$ and the right-handed neutrinos.  
The right-handed neutrinos would later decay to leptons and other particles, and we are left with a relic density of $\chi$ 
which  is to be compared with the DM relic density found in cosmological observations.

We assume that  coupling  $y_{DS}$ is very small, typically  at order $10^{-12}$.
If $y_{DS}$ is not small, the  fermion DM $\chi$ can be produced thermally\cite{macias2016realistic,escudero2017sterile,blennow2019neutrino}
which is not the case we consider here.
For such a small value of $y_{DS}$, the fermion $\chi$ is never in thermal equilibrium.
However, the annihilation of $\phi$  and $N$ or the decay of $\phi$ in the thermal bath can  also produce $\chi$ and contribute to the relic density of $\chi$, 
through the so-called freeze-in mechanism.
So the total relic density of DM $\chi$ consists of contributions from two parts. 
One part is from the decay of $\phi$ and the other is from freeze-in production of $\chi$.
We can write the total relic abundance of DM as 
\begin{equation}
\rho_{\chi}=m_{\chi} s_{0} Y_{DM}
\end{equation}
which is usually written in terms of the density parameter $\Omega$ 
and the critical mass density $ \rho_{\text{c}} $
\begin{equation}
\label{relicdensity}
\Omega_{\mathrm{DM}} h^{2} = \frac{\rho_{\chi}} { \rho_{\text {c }} / h^{2}}
\end{equation}
where $s_0 = 2891.2$ $\text{cm}^{-3}$ is the entropy density of today 
and $\rho_c$ is the critical density of the universe with $ \rho_{\text{c}}/h^2= 1.05 \times 10^{-5}$ $\text{GeV}/\text{cm}^3$.
Here we define $Y_i = n_i/s$ where $n_i$ is the corresponding number density.
$Y_{DM}$ can be expressed as
\begin{equation}
Y_{DM} = Y_{\chi} + Y_{\phi}, \label{YDM}
\end{equation}
 where $Y_{\chi}$ denotes the contribution from the freeze-in production and $Y_{\phi}$ the contribution from the decay of $\phi$.

\begin{figure}[!t]
	\centering
	\subfigure[\label{sketch1}]
	{\includegraphics[width=.486\textwidth]{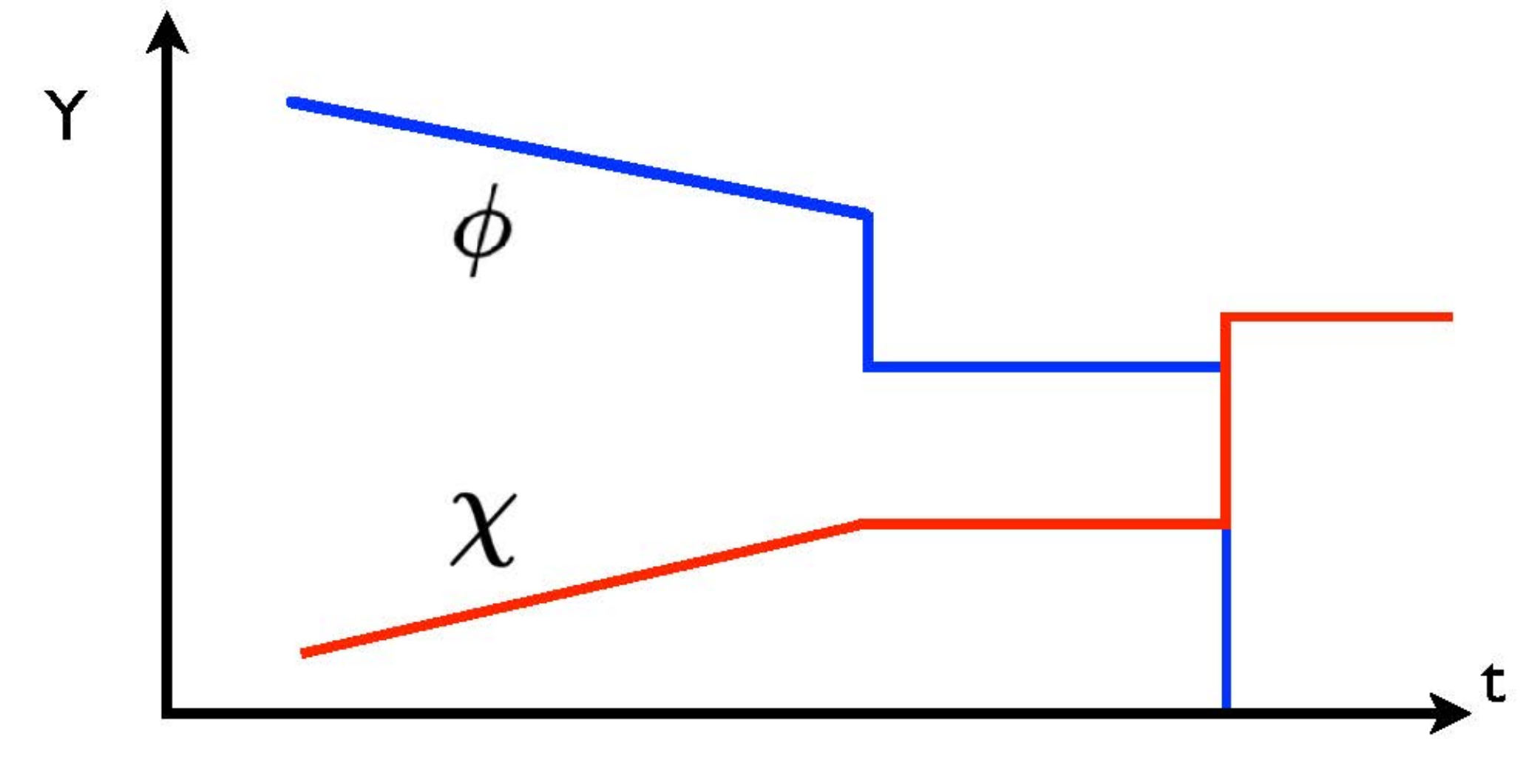}}
	\subfigure[\label{sketch2}]
	{\includegraphics[width=.486\textwidth]{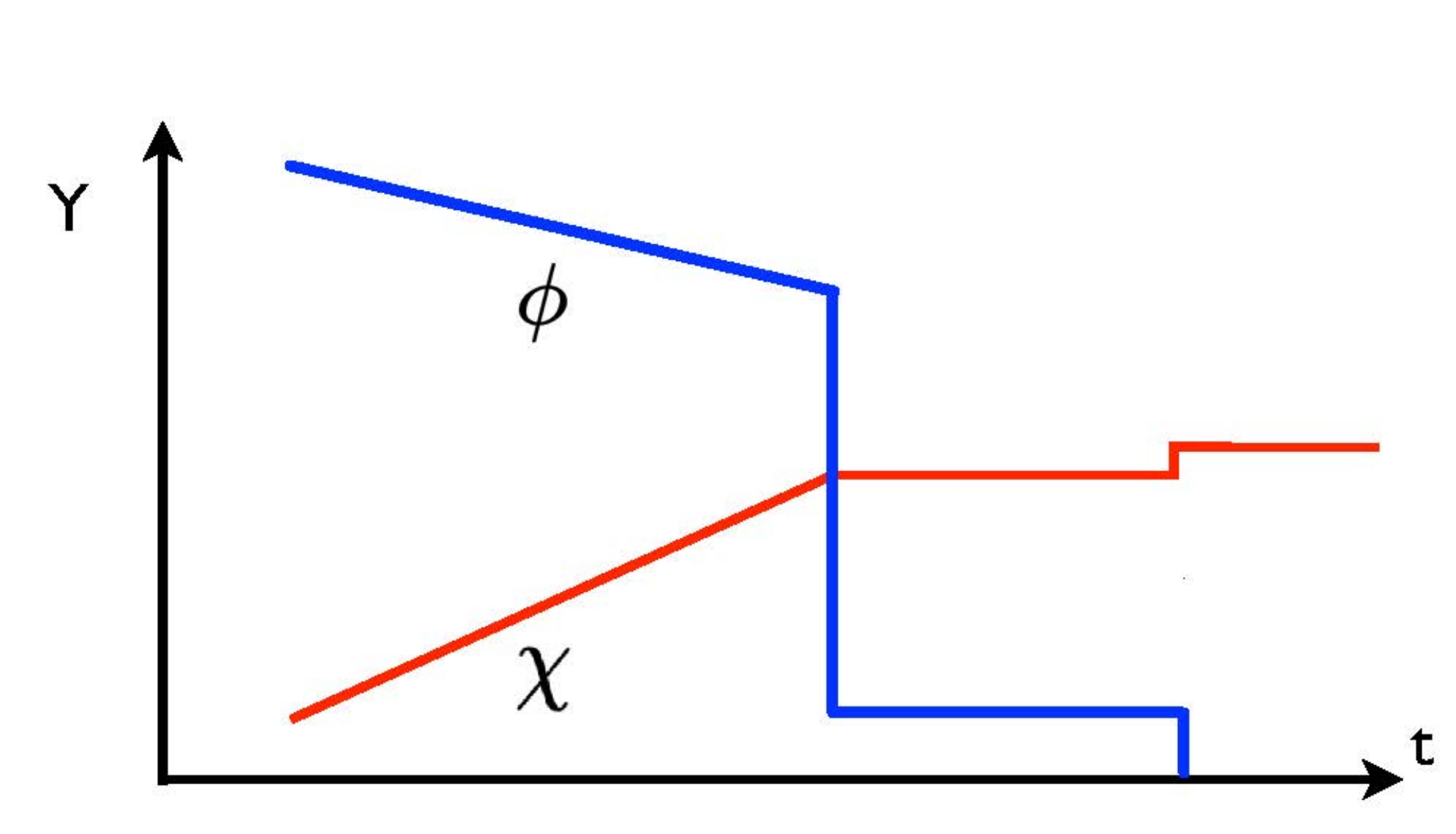}}
	\caption{The diagrammatic sketch of the two typical scenarios involving the production of DM relic density. 
	    (a)  The dominant contribution to the relic abundance of DM $\chi$ is from the decay of $\phi$ after freeze-out.
		(b)  The dominant contribution to the relic abundance of DM $\chi$ is generated by the freeze-in mechanism.
	}
	\label{fig:sketch}
\end{figure}

 In  Fig.~\ref{fig:sketch}, we show a diagrammatic sketch of the production of $\chi$.
Fig.~\ref{sketch1} shows that the density of $\chi$ gradually accumulates by the freeze-in, and then suddenly increases to a significant amount
after the $\phi$ freezes out and then decays to $\chi$. This is the case that the DM relic density is dominated by the decay of $\phi$ after freeze-out.
Fig.~\ref{sketch2} shows the opposite case. The density of $\chi$ gradually accumulates to a significant amount by the freeze-in mechanism,
and then suddenly increases by a small amount as  the $\phi$ freezes out and then decays to $\chi$. 
This is the case that the DM relic density is dominated by the freeze-in production of $\chi$.
The former case corresponds to the case with smaller $y_{DS}$ so the production process is weaker.
The later case corresponds to the case with larger $y_{DS}$ so the production process is stronger.
More details will be presented later

Determining the relic abundance of the DM  requires solving the Boltzmann equations
which describe the time evolution of the number densities of particles in the expanding universe. 
In the case under consideration, $\chi$ is never in thermal equilibrium and we can neglect
the contribution by the annihilation of $\chi$ in the Boltzmann equations. 
The Boltzmann equations for the two dark sector particles are given as
\begin{eqnarray}
\frac{d n_{\phi}}{d t} &&=-3 H n_{\phi}-\langle\sigma v\rangle_{\phi \phi \rightarrow N N}\left(n_{\phi}^{2}-(n_{\phi}^{e q})^2\right)-\langle\sigma v\rangle_{\phi \phi \rightarrow SM}\left(n_{\phi}^{2}-(n_{\phi}^{e q})^2\right)-\Gamma_{\phi} n_{\phi}, \label{Bolt1}\\
\frac{d n_{\chi}}{d t} &&=-3 H n_{\chi}+\langle\sigma v\rangle_{NN \rightarrow \chi \chi }n_{N}^2+ \langle\sigma v\rangle_{h N \rightarrow \phi \chi } n_{N} n_{h} + \langle\sigma v\rangle_{h \nu \rightarrow \phi \chi } n_{\nu} n_{h} +\Gamma_{\phi} n_{\phi},  \label{Bolt2}
\end{eqnarray}
where $H=1.66 \sqrt{g^{\rho}_{*}(T)} T^{2}/M_{\mathrm{Planck}}$  and  $s=2 \pi^{2} g^{s}_{*}(T) T^{3}/45$. 
$M_{pl}$ is the Planck mass, and  $g^\rho _{*}$ 
and  $g^s_{*}$ are the effective number of degrees of freedom of the relativistic species in the thermal bath
for energy density and entropy density  respectively.

\begin{figure}[!t]
	\centering
	\subfigure[\label{phiphi_NN}]
	{\includegraphics[width=0.31\textwidth]{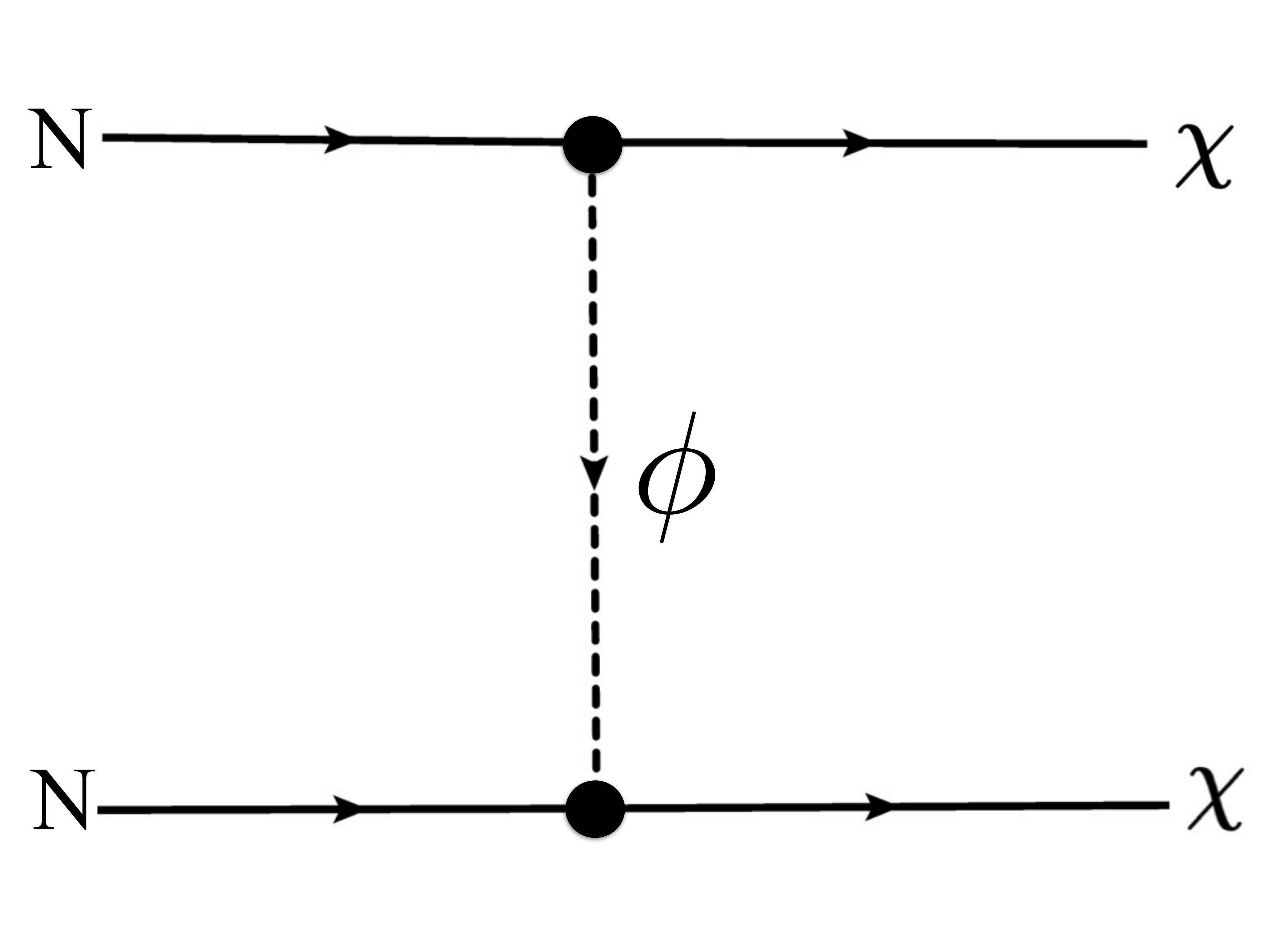}} 
	\subfigure[\label{hNscattering}]
	{\includegraphics[width=0.31\textwidth]{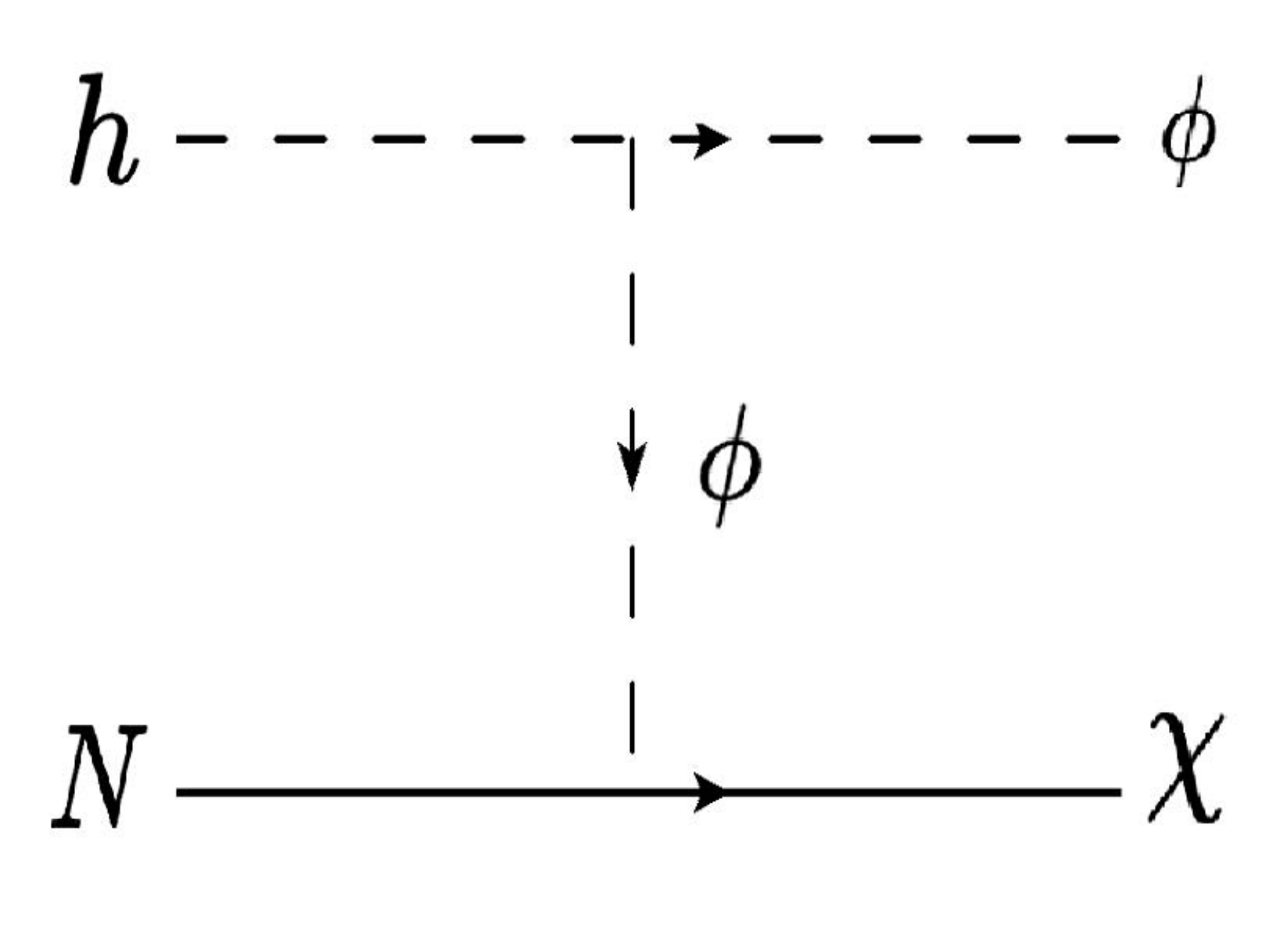}}
	\subfigure[\label{hvscattering}]
	 {\includegraphics[width=0.31\textwidth]{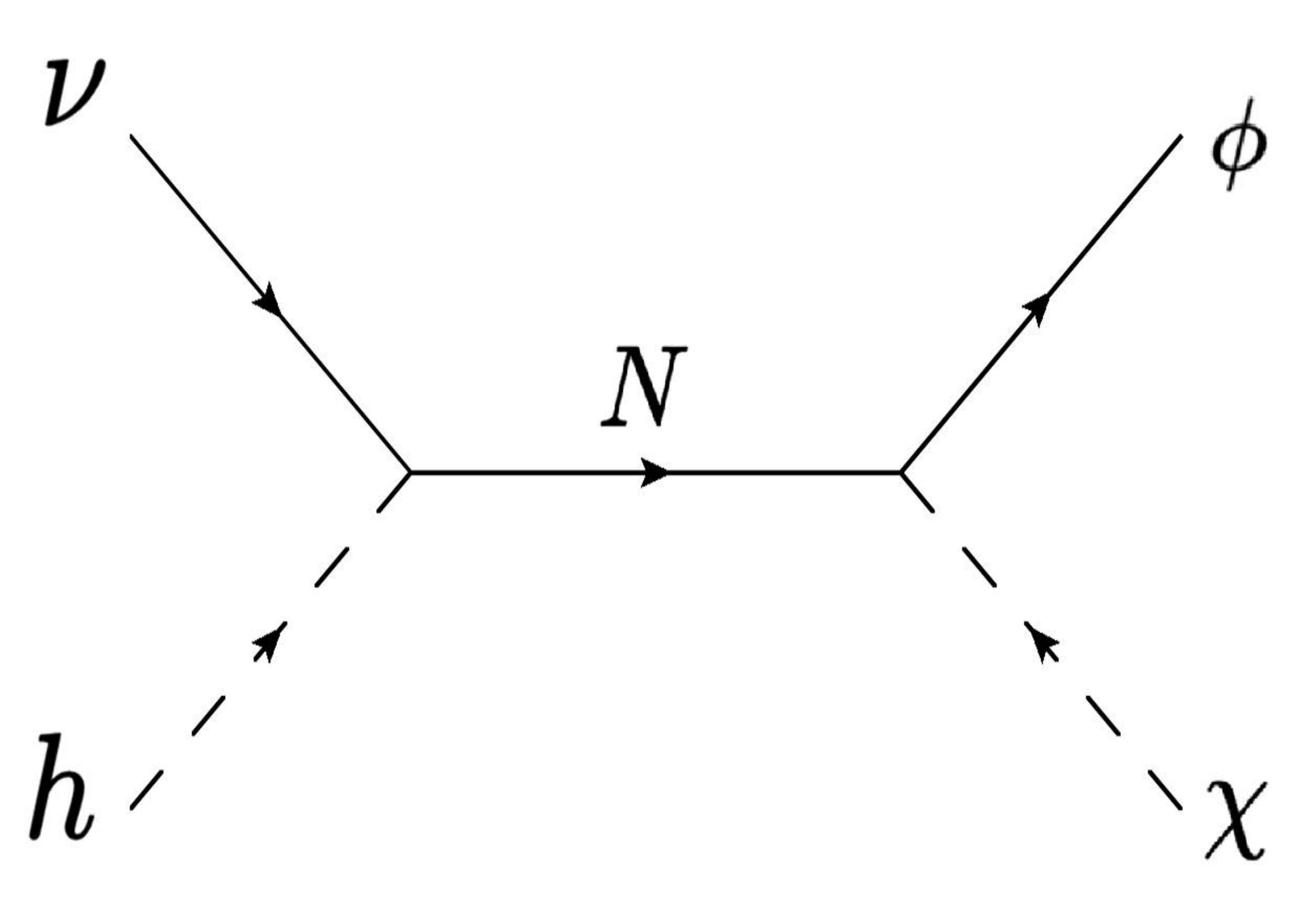}}
	 \subfigure[\label{chichi_NN}]
	 {\includegraphics[width=0.31\textwidth]{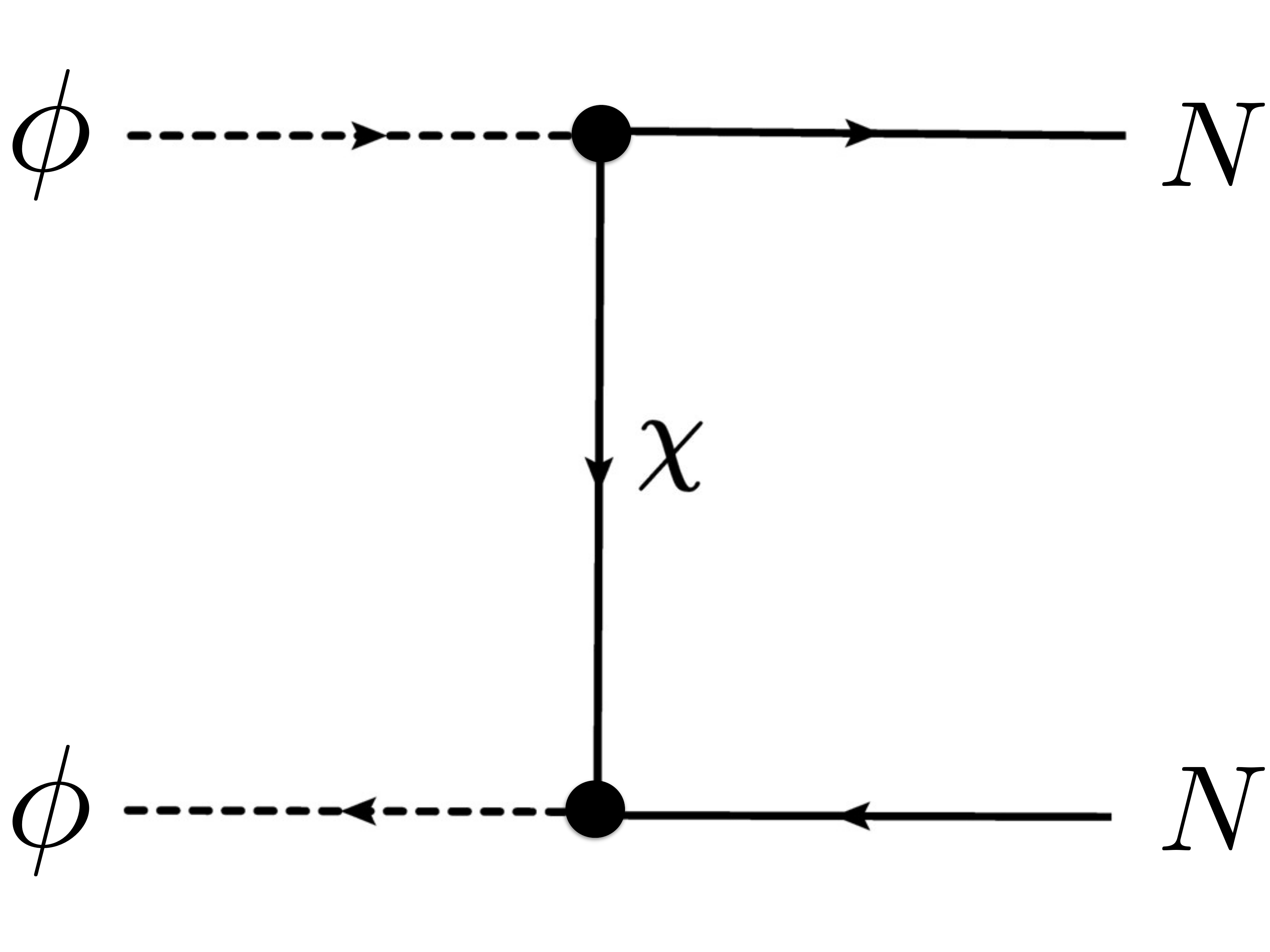}}
	\subfigure[\label{phiphi_SMS}]
	{\includegraphics[width=0.31\textwidth]{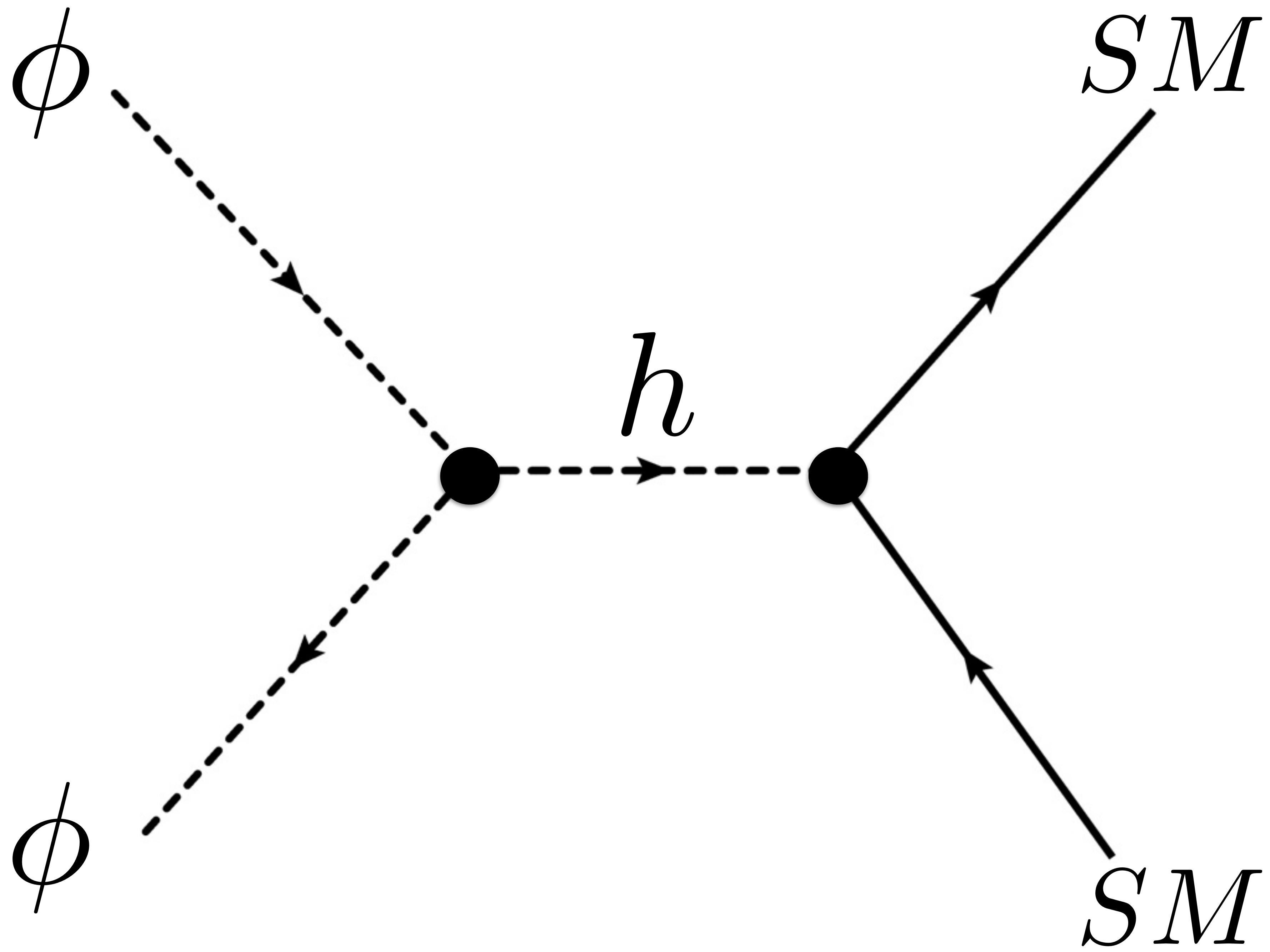}}
	\caption{Relevant annihilation and scattering channels.} 
	\label{fig:Anndiagrams}
\end{figure}

The second terms in both equations (\ref{Bolt1}) and (\ref{Bolt2})  correspond to the annihilation processes involving two sterile neutrinos.
 The third term in Eq.~\eqref{Bolt1} refers to the processes of annihilation into the SM particles,
 e.g., $\phi \phi \rightarrow h h $, $\phi \phi \rightarrow W^{+} W^{-}$ and $\phi \phi \rightarrow t \bar{t}$ etc.
The third term in Eq.~\eqref{Bolt2} corresponds to a t-channel process of Higgs and right handed neutrinos 
and the fourth term  corresponds to the Higgs and neutrino annihilation process originating from the 
Yukawa interaction of the left and right handed neutrinos.
 The relevant Feynman diagrams are shown in Fig.~\ref{fig:Anndiagrams} and the expressions of corresponding cross sections are given in 
 Appendix \ref{App:crosssections}. 
 
 The last terms in Eqs.  (\ref{Bolt1}) and (\ref{Bolt2}) correspond to the decay of $\phi$ into $\chi$ and the right-handed neutrinos.
 It is governed by 
 the two body decay rate $\Gamma_\phi=\Gamma_{\phi \rightarrow \chi N}^{2-\mathrm{body}}$ of  scalar particle $\phi$  which can be expressed as
\begin{equation}
\label{decayrateofphi}
\Gamma_{\phi}=\frac{y_{\mathrm{DS}}^{2} m_{\phi}}{16 \pi}\left(1-\frac{m_{\chi}^{2}}{m_{\phi}^{2}}-\frac{m_{N}^{2}}{m_{\phi}^{2}}\right) \lambda^{\frac{1}{2}} \left(1, \frac{m^2_{\chi}}{m^2_{\phi}}, \frac{m^2_{N}}{m^2_{\phi}}\right),
\end{equation}
where $\lambda$ is the Kallen function given as
\begin{equation}
\lambda(x, y, z) \equiv x^{2}+y^{2}+z^{2}-2 x y-2 x z-2 y z.
\end{equation}
 
 To solve Eq.~\eqref{Bolt1},
we notice that the cross section of the annihilation processes $\phi\phi \to NN$ 
is proportional to $y^{4}_{DS}$,  so it's highly suppressed in our scenario. 
The decay rate is also proportional to $y^{2}_{DS}$.
So the first and the third terms in Eq.~\eqref{Bolt1} are the  dominant terms and Eq.~\eqref{Bolt1} can be solved by the freeze-out approximation\cite{griest1991three,kolb1981early}. We obtain the relic density parameter of $\phi$ as
\begin{equation}
\Omega_{\phi} h^{2} \approx \frac{1.07 \times 10^{9} \mathrm{GeV}^{-1}}{M_{\mathrm{pl}}} \frac{x_{f}}{\sqrt{g^s_{\star}}} \frac{1}{a+3 b / x_{f}},
\end{equation} 
and 
\begin{equation}
Y_{\phi}  \approx \frac{3.89 x_f}{\sqrt{g^s_{\star}} M_{\mathrm{pl}} m_{\phi}(a+3 b / x_{f})} , \label{Yphi}
\end{equation} 
where $Y \equiv n / s$ and the parameters $a$ and $b$ are parameters appearing in the annihilation cross section when decomposed as $\langle\sigma v\rangle=a+b v^{2}$. The freeze-out temperature $x_f = m_{\phi}/T_{f}$ is determined numerically by
\begin{equation}
x_{f}=\ln \frac{0.038 g_{\phi} M_{\mathrm{Pl}} m_{\phi}\langle\sigma v\rangle}{g_{*}^{1 / 2} x_{f}^{1 / 2}}.
\end{equation}
 After freeze-out, $\phi$ would totally convert through decay into $\chi$ and the right-handed neutrinos. 
So the $Y_\phi$ obtained in Eq. \eqref{Yphi} gives one of the two contributions shown in Eq. \eqref{YDM}.

To solve Eq.~\eqref{Bolt2}, 
we notice again that the cross section of the annihilation process $N N \rightarrow \chi \chi$ is also proportional to $y^4_{DS}$ and  is highly suppressed. 
The $h N \rightarrow \phi \chi$ and $h v \rightarrow \phi \chi$ processes are proportional to 
$\lambda^2_{H \phi} y^2_{DS}$ and $y^2_{\nu} y^2_{DS}$ respectively. 
Here $y_{\nu} = \left( U^{\dagger} Y \right) / \sqrt{2}$, and the matrix U is the PMNS matrix in neutrino mixing.
 For $\lambda_{H \phi}$ and the magnitude of $y_{\nu}$ less than 1, these two processes are also suppressed so they are subdominant.
 The dominant contribution of producing $\chi$ DM comes from the decay of the heavier $\phi$ in thermal equilibrium.
 The scalar $\phi$ is in the thermal equilibrium with the SM particles for temperature above the freeze-out temperature $T_f$.
 So we can take $n_{\phi} \approx n^{eq}_{\phi}$ and Eq.~\eqref{Bolt2} can be solved approximately\cite{hall2010freeze}. 
 We can get
\begin{equation}
\label{Ychi}
Y^{}_{\chi} \approx \frac{45}{(1.66) 4 \pi^{4}} \frac{g_{\phi} M_{P l} \left( \Gamma_{\phi \rightarrow \chi N}+\Gamma_{\phi \rightarrow \bar{\chi} N} \right) }{m_{\phi}^{2} g_{*}^{s} \sqrt{g_{*}^{\rho}}} \int_{x_{f}}^{x_{\max}} K_{1}(x) x^{3} d x,
\end{equation}
where $x \equiv m / T$ and $K_1$ is the first modified Bessel Function of the second kind.

 So $Y_{DM}$ can be obtained by summing Eq.~\eqref{Ychi} and Eq.~\eqref{Yphi}.
Thus we can obtain the total relic abundance by Eq.~\eqref{relicdensity}.
 The expression for $\Omega_{\mathrm{DM}}$ can now be compared to the experimental value obtained 
by the Plank Collaboration at 68\% CL\cite{aghanim2018planck}:
\begin{equation}
\Omega_{Planck} h^{2}=0.1192 \pm 0.0010
\end{equation}

We note that one of the product of the  $\phi$ decay is the right-handed neutrino which would decay to the SM particles and increase the entropy.
We assume the couplings of right-handed neutrinos with the SM particles are not very small so that they decay rapidly into SM particles.
For N much heavier than W and Z boson, N would decay mainly through the two body processes $N \rightarrow l^{\pm} W^{\mp}$ and $N \rightarrow \nu(\bar{\nu}) Z$.
There is an additional channel $N \rightarrow \nu(\bar{\nu}) H$ for the case $m_N \gg m_h$. 
 The decay rate of N for each channel can be found as\cite{Atre:2009rg,He:2009mv,Banerjee:2015gca,delAguila:2008cj,Pilaftsis:1991ug}
\begin{equation}
\label{Ntwobodydecay}
\Gamma(N \rightarrow V f) \approx \frac{g^{2}}{64 \pi m_{W, Z, h}^{2}}\left|R_{l N}\right|^{2} m_{N}^{3}, 
\end{equation}
where $R_{l N}$ is the mixing matrix between the heavy and active neutrinos. 
Further two body decay of Z and W bosons would give partial decay rates of sterile neutrino decaying to three body final states.
In the case $m_N<m_W$,  N would decay mainly to three fermions through off-shell W and Z. 
The decay rate of N for each channel can be found as\cite{Pilaftsis:1991ug,Liao:2017jiz}: 
\begin{equation}
\label{Nthreebodydecay}
\Gamma(N \rightarrow 3 f) \approx \frac{G_{F}^{2}}{192 \pi^{3}}\left|R_{l N}\right|^{2} m_{N}^{5}. 
\end{equation}
In fact, the decay rate of sterile neutrino decaying to three final fermions through interaction with Z and W bosons 
can be written in a more general way\cite{Liao:2017jiz}.
This formalism includes both the effects of the on-shell and the off-shell Z and W bosons and can be used in general parameter space,
not just for $m_N<m_W$. For $m_N > m_Z$ or  $m_N < m_W$, this formalism gives results consistent with results calculated using
Eqs. (\ref{Ntwobodydecay}) and (\ref{Nthreebodydecay})\cite{Liao:2017jiz}.  It can also be used in the range of parameter
space $m_N \sim m_W$ or $m_Z$. 
More details of this formalism for each decay process can be found in the Appendix~\ref{App:Ndecaywidths}.

We calculate the lifetime of the heavy sterile neutrinos by taking into account all the three body final states and the results are shown in Fig.~\ref{lifetimeofNR}. 
In this calculation, we use the formulas in Appendix.~\ref{App:Ndecaywidths} which include both the effects of exchanging on-shell 
and off-shell Z and W bosons by including the width of W and Z in the propagators. 
In the parameter space shown in Fig.~\ref{lifetimeofNR},
the lifetime of the heavy sterile neutrinos would be much shorter than the lifetime of the dark sector particle $\phi$ which we consider in this model,
as long as the magnitude of the mixing $R_{l N}$ is well above the order of $10^{-10}$. 
Since $R_{lN}$ is basically a free parameter which does not affect Eqs. (\ref{Yphi}) and (\ref{Ychi}), 
this conclusion works for both scenarios described in Fig. \ref{fig:sketch}.
So the entropy increase happens shortly after the decay of $\phi$. 
We can simply choose the time of entropy  increase as the time of $\phi$ decay which we assume  to happen at MeV temperature scale.
The entropy increase can be estimated as
\begin{equation}
\frac{\Delta s}{s} \approx \frac{n_{\phi}(m_{\phi}-m_{\chi})}{s T} \approx \frac{Y_{\phi} (m_{\phi}-m_{\chi})}{T} \approx 10^{-4} \ll 1
\end{equation}
 for $T\sim $MeV and the time at about 1 second. We find that the entropy increase from the decay can be neglected.

\begin{figure}[!t]
	\centering
	\includegraphics[width=0.5\textwidth]{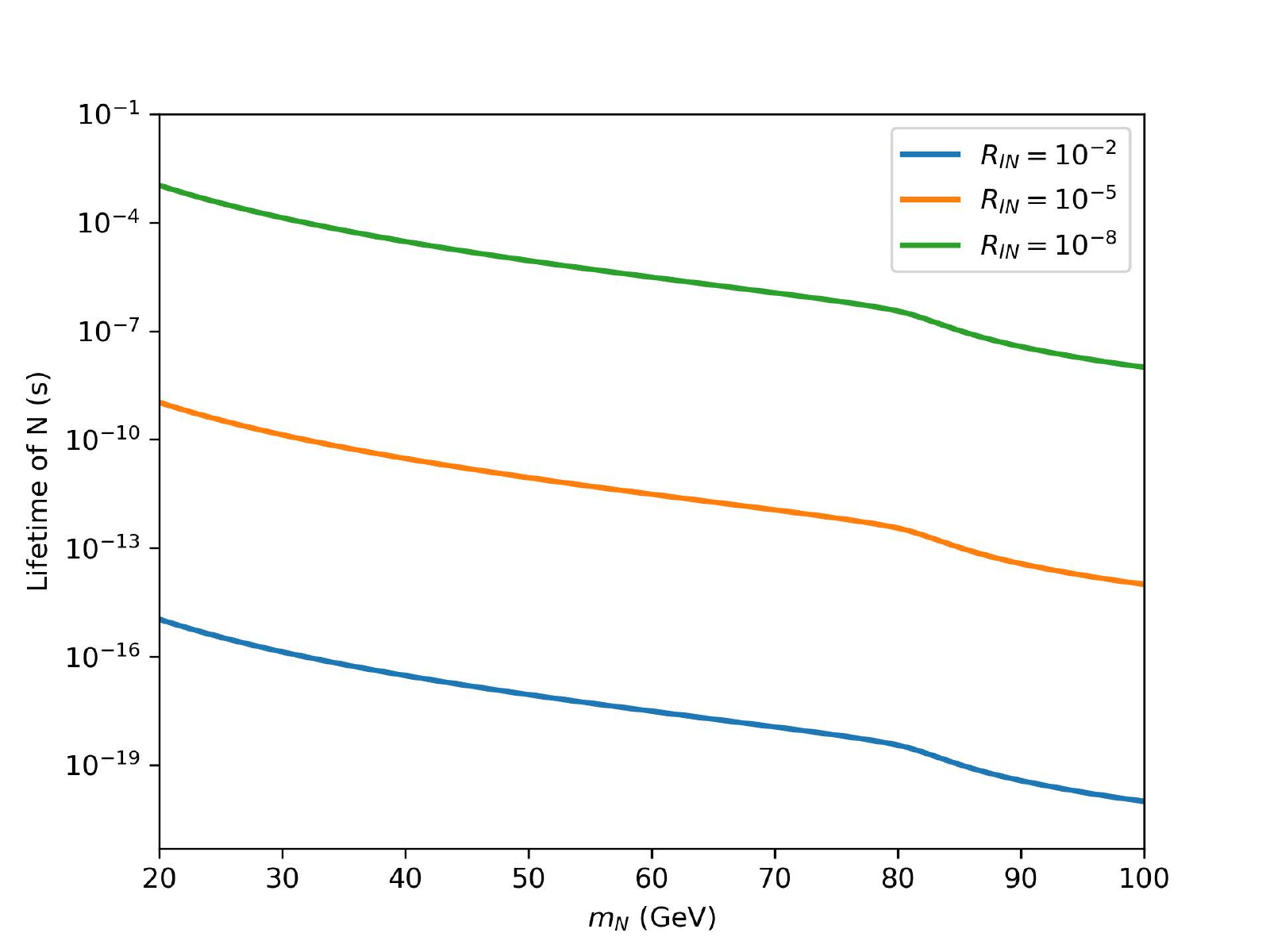}  
	\caption{ The lifetime of the heavy sterile neutrinos for different mixing $R_{lN}$.  } 
	\label{lifetimeofNR}
\end{figure}

\begin{figure}[!t]
	\centering
	\subfigure[\label{fig:BBNlimit1}]
	{\includegraphics[width=.486\textwidth]{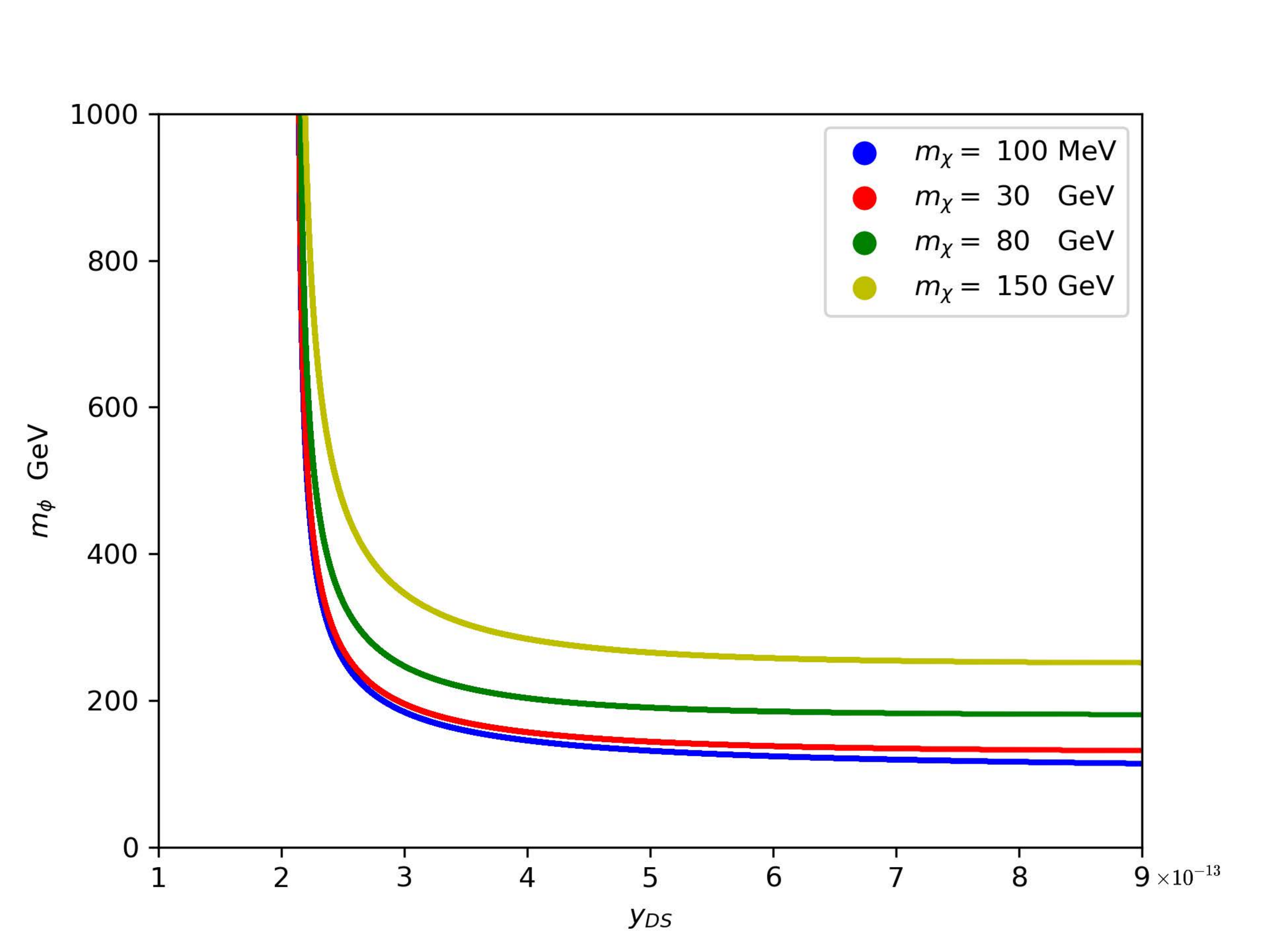}} 
	\subfigure[\label{fig:BBNlimit2}]
	{\includegraphics[width=.486\textwidth]{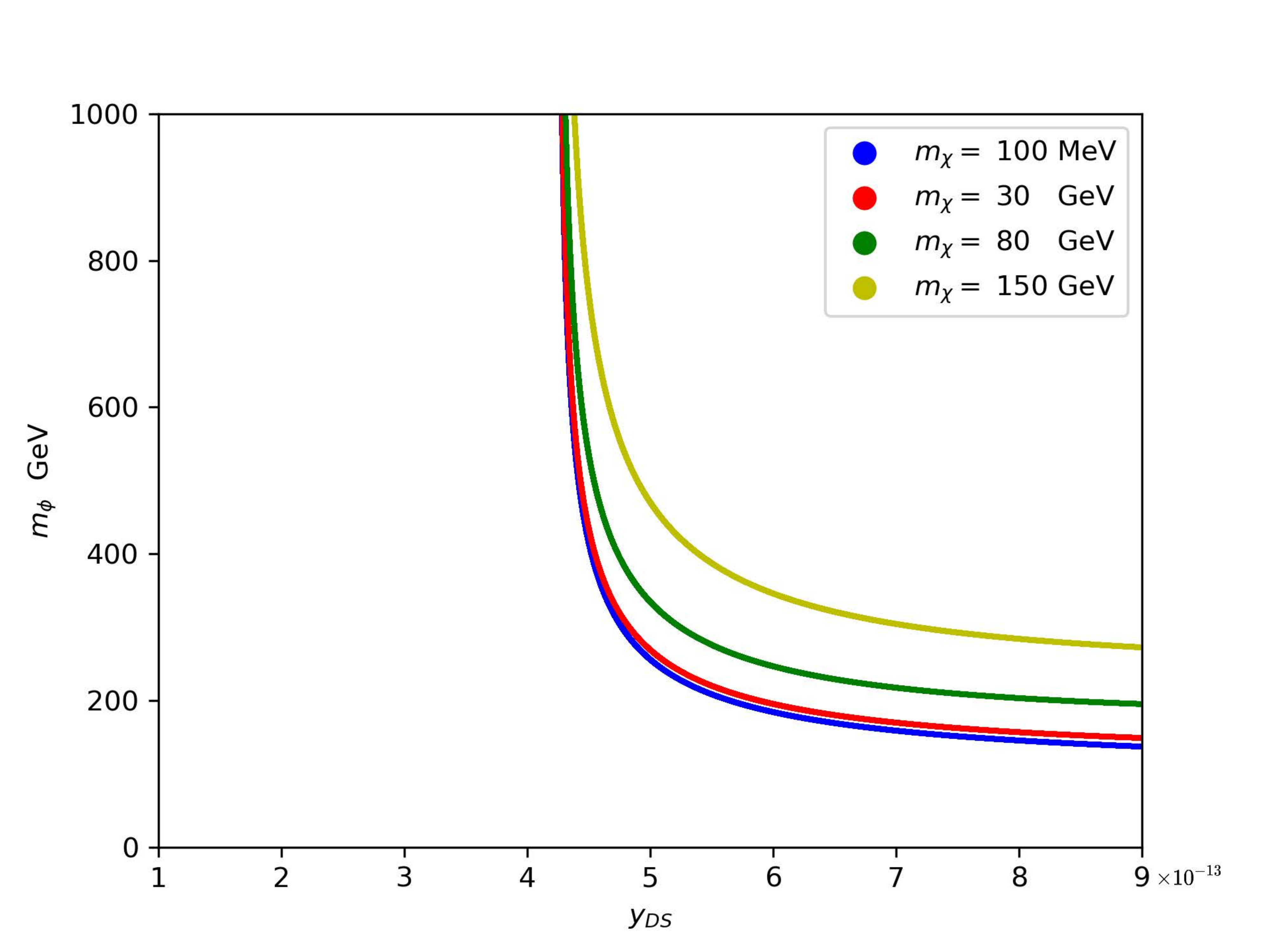}}
	\subfigure[\label{fig:BBNlimit5}]
	{\includegraphics[width=.486\textwidth]{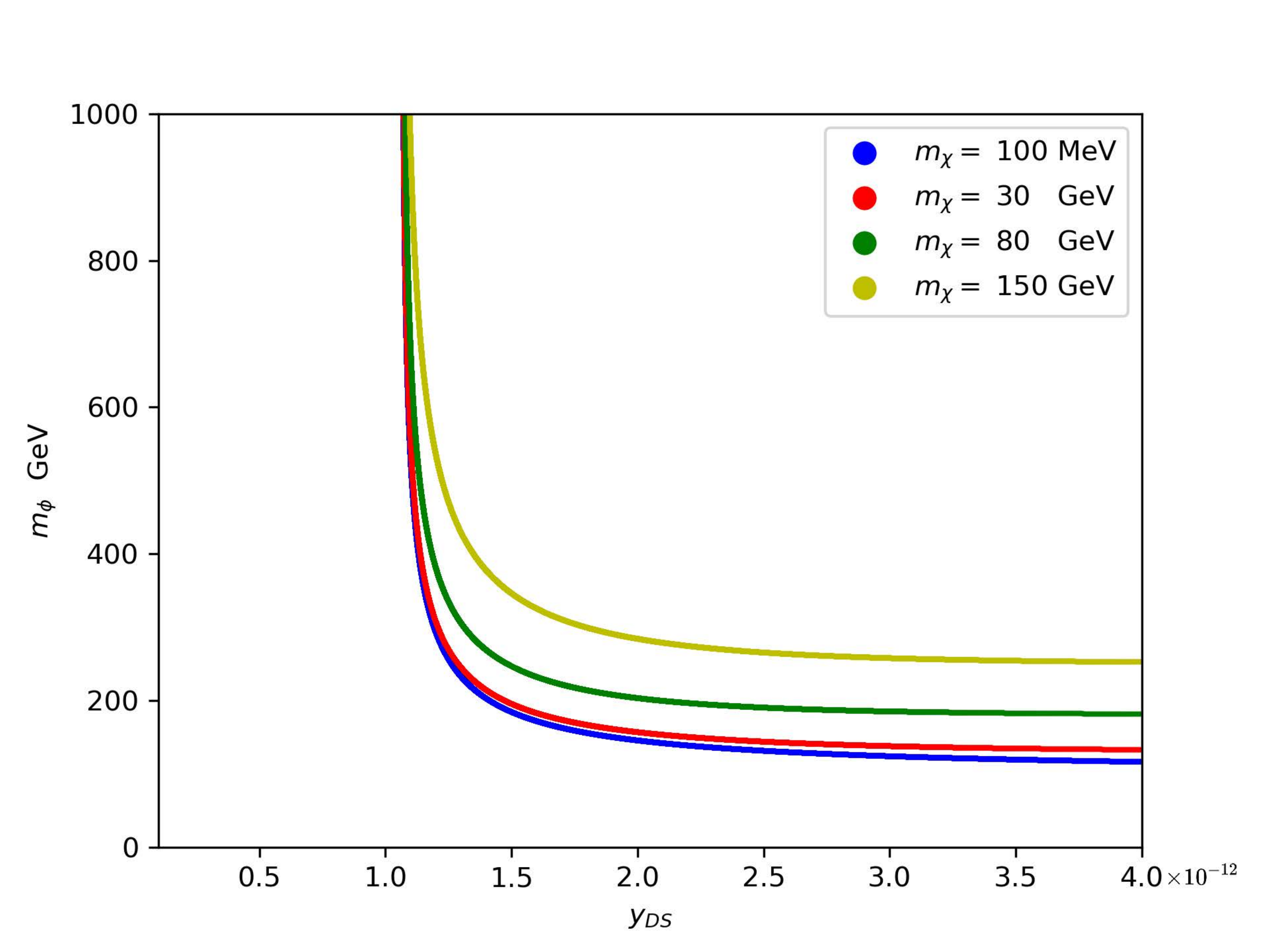}}
	\subfigure[\label{fig:BBNlimit10}]
	{\includegraphics[width=.486\textwidth]{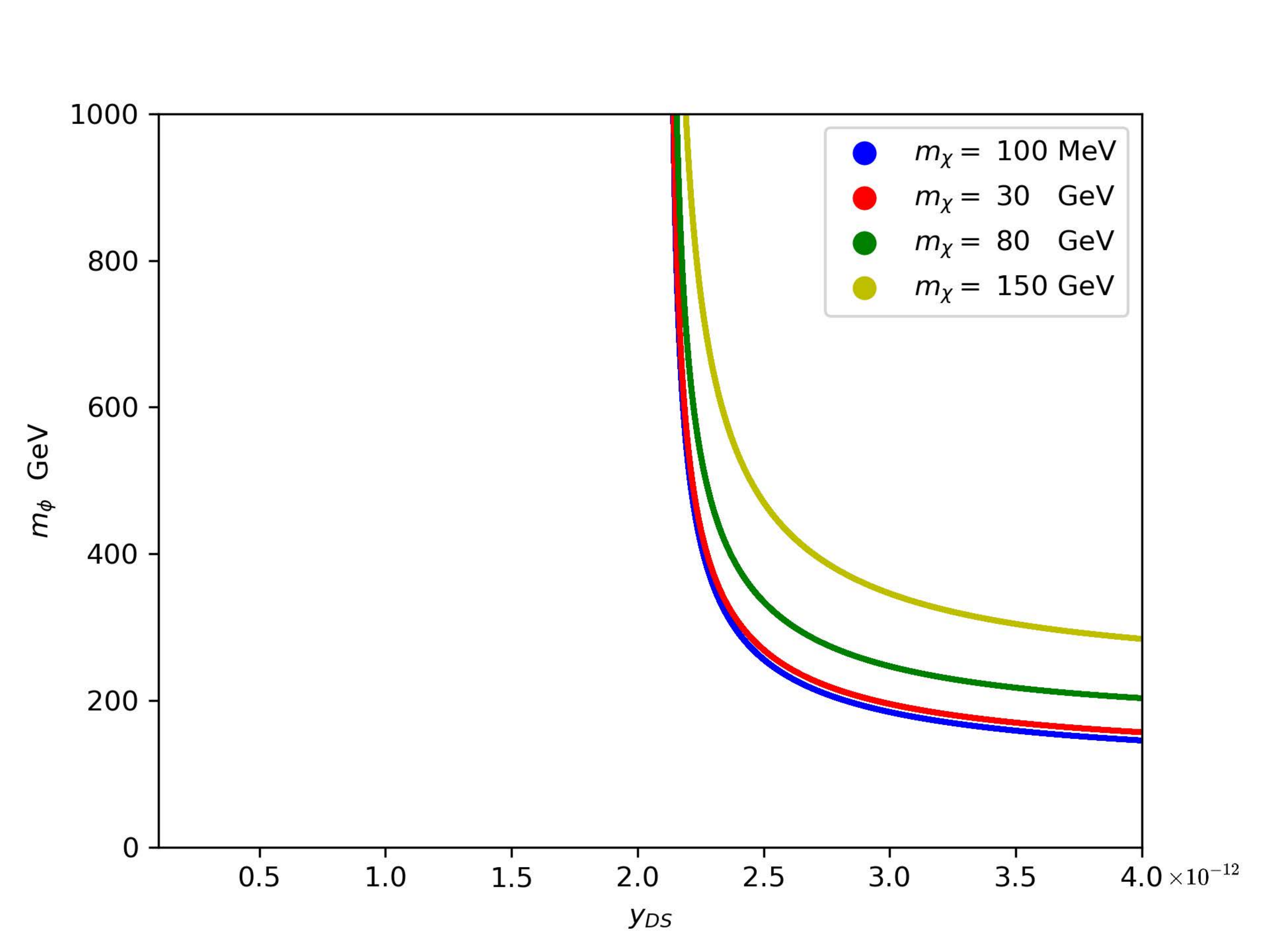}}
	\caption{Constraint on parameter space from lifetime consideration. 
	 Parameter space above the lines are allowed.
	 (a) T = 1 MeV, t $\approx$ 0.7 s. (b)T = 2 MeV, t $\approx$ 0.2 s. (c) T = 5 MeV, t $\approx$ 0.03 s. (d)T = 1 MeV, t $\approx$ 0.007 s.} 
	\label{fig:BBNlimit}
\end{figure}

We choose the $\phi$ to decay at MeV scale, in particular before the temperature drops down to BBN scale.
This is because the energy release given by $\phi$ decay would give rise to significant amount of relativistic particles
which may have potential effect on the BBN.
To avoid possible effect on the standard BBN scenario, 
 we conservatively demand that $\phi$ should decay before the temperature drops down to MeV scale. 
 For the radiation-dominated epoch, $t=0.301 (g_{*}^{\rho})^ {-1 / 2} m_{P l}/T^{2}$ where  $g_{*}^{\rho} = 10.75$ 
 is the effective numbers of degrees of freedom of the relativistic species in the thermal bath 
 when 100 $\mathrm{MeV} \gtrsim T \gtrsim$ 1 $ \mathrm{MeV}$\cite{kolb1981early}. 
 So  decay at MeV temperature scale corresponds to the lifetime of $\phi$ about 0.01 \rule[+3pt]{0.3cm}{0.05em} 1 seconds. 
 In Fig.~\ref{fig:BBNlimit}, we show the effect of this constraint on the allowed parameter space.
 We choose four benchmark points of the temperature to be 1, 2, 5 and 10 MeV which correspond to the time about 0.7, 0.2, 0.03 and 0.007 second separately. 
 The allowed parameter spaces for $\phi$ decay before the temperature dropping below theses benchmark points are shown in Fig.~\ref{fig:BBNlimit}. 
 In the calculation we choose $m_{N} = 100$ GeV. 
 The lines in the figure represent values of  parameters that make the lifetime of $m_{\phi}$ exactly the time of the chosen benchmark points.
 The regions above the line are regions with shorter lifetime of $m_{\phi}$. 
 Different colors represent different values of $m_{\chi}$. 
 In the figure, we can find low bounds on $y_{DS}$. 
 We can see that $y_{DS}$ is typically constrained to be above the order around $10^{-13}-10^{-12}$.
For $m_{\chi} < 100$ MeV, the  constraint on $y_{DS}$ is similar to the case $m_{\chi} = 100$ MeV. 
 We can see that this consideration does not give a strong constraint on the parameter space which gives the right DM relic abundance.

\section{Numerical results}
\label{sec:numerical}
We use the public code \textbf{MicrOmegas}\cite{belanger2014micromegas_3,belanger2018micromegas5} and 
\textbf{CALCHEP}\cite{belyaev2013calchep} to solve the Boltzmann equation and calculate relic abundance. 
In numerical calculation, all the contributions from $2 \rightarrow 2$ process have been taken into account. 
The model implementation for \textbf{MicrOmegas} was done using \textbf{FeynRules}\cite{alloul2014feynrules} package. 
There are five free parameters in the model,  i .e. $m_{\phi}$,$m_{\chi}$,$m_{N}$,$y_{DS}$ and $\lambda_{H \phi}$.
We constrain all these parameters using the observed DM relic abundance.

\begin{figure}[!t]
	\centering
	\includegraphics[width=0.5\textwidth]{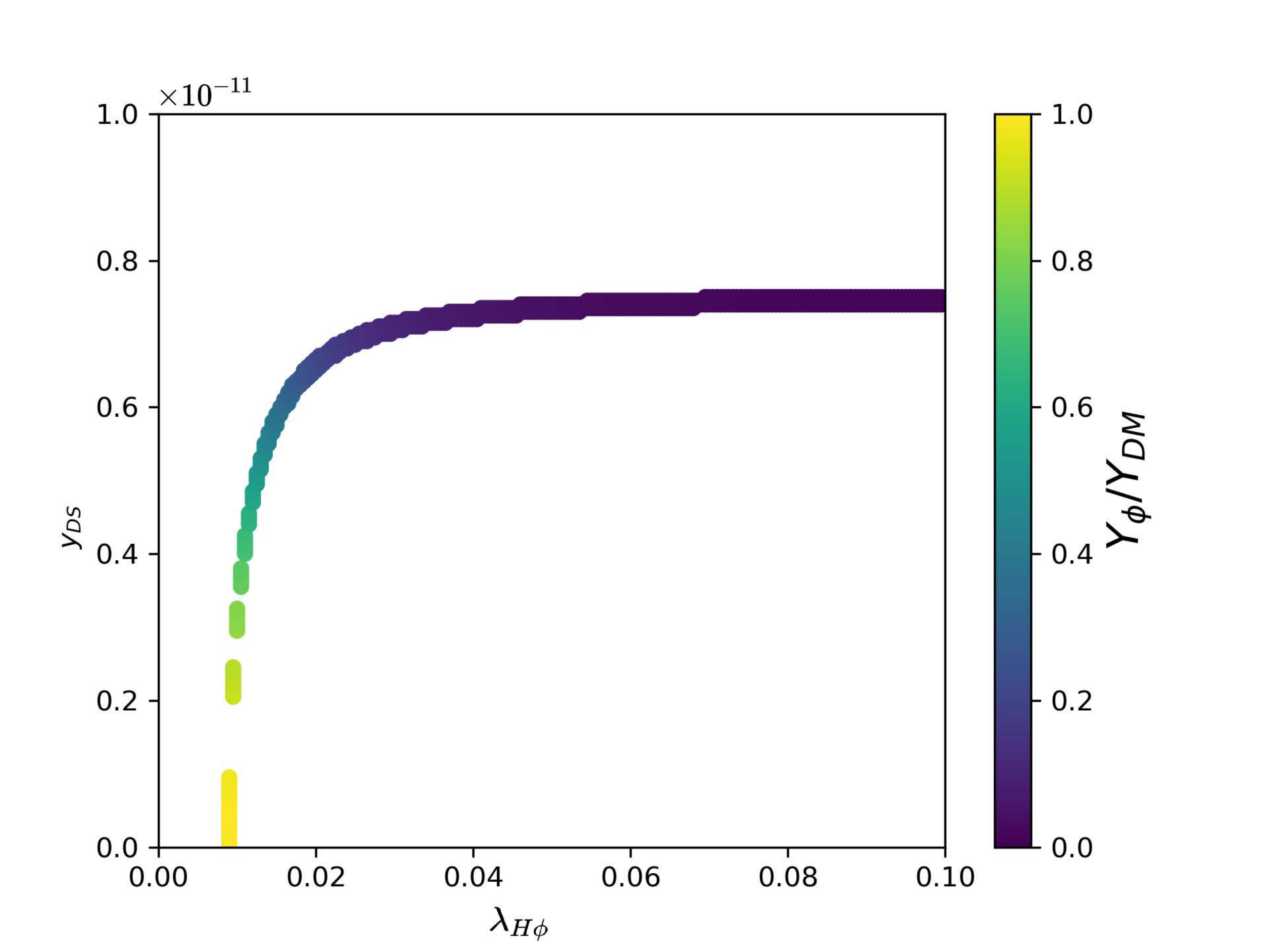}  
	\caption{The allowed parameter space of $\lambda_{H \phi}$ and $y_{DS}$.
	$M_{N} = 100$ GeV, $m_{\phi} = 500$ GeV and $m_{\chi} = 10$ GeV.
	Different colors correspond to different values of the ratio of contribution from the decay of $\phi$ after freeze-out 
	to the total $\chi$ DM relic density.} 
	\label{yDS_lambda}
\end{figure}

\begin{figure}[!t]
	\centering
	\subfigure[\label{Mphilambda}]
	{\includegraphics[width=.486\textwidth]{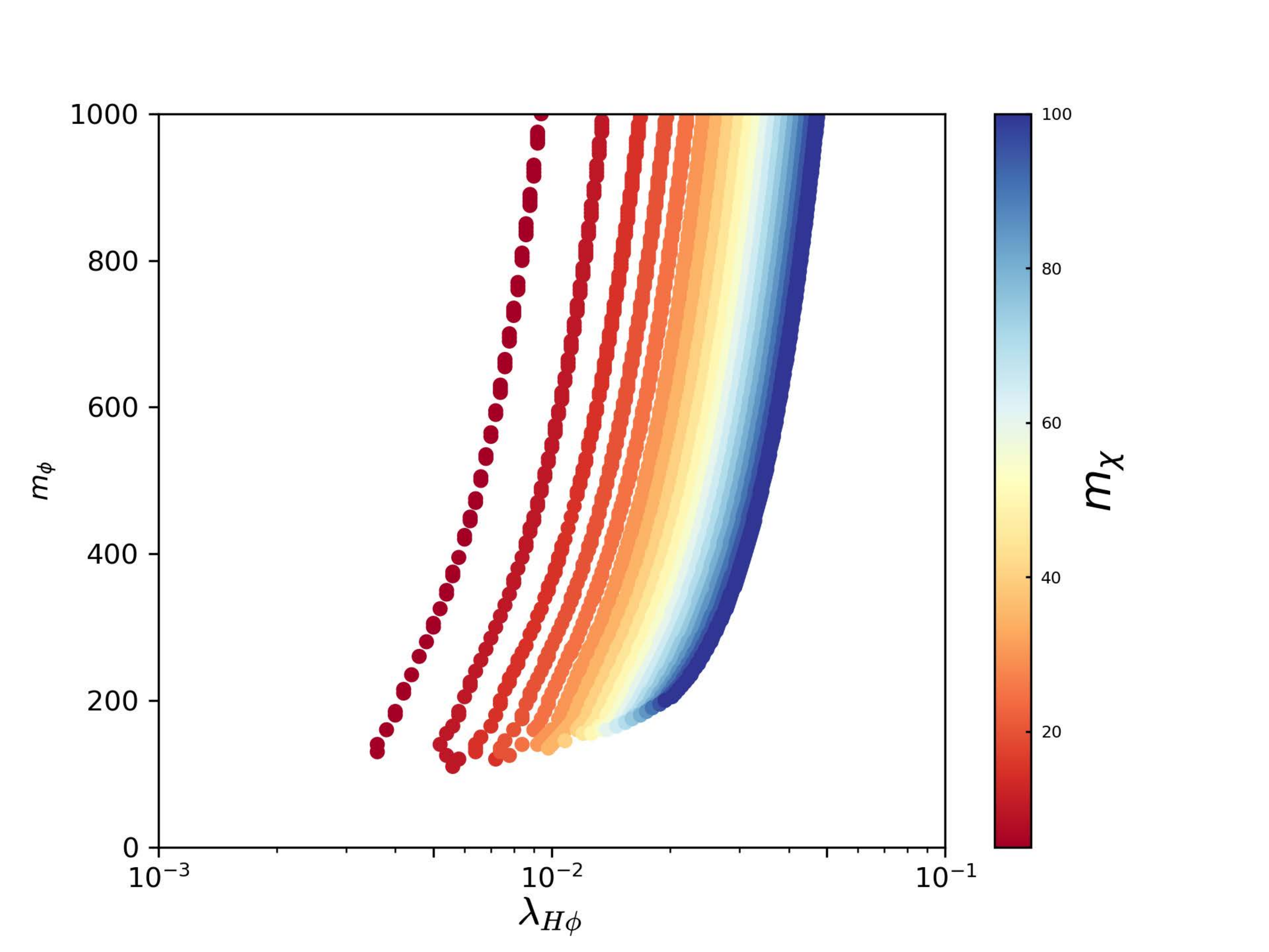}}
	\subfigure[\label{MphiYds}]
	{\includegraphics[width=.486\textwidth]{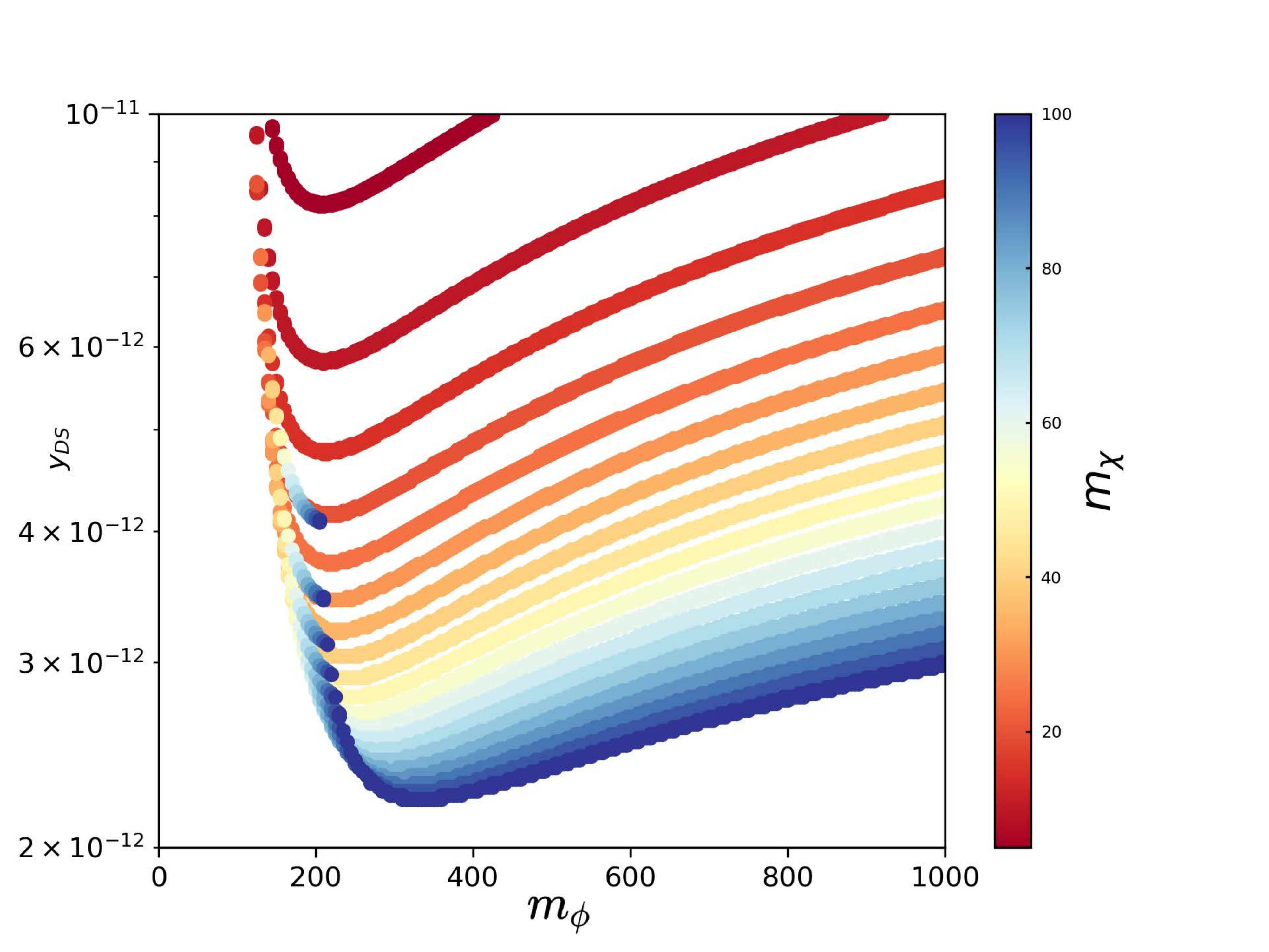}}
	\caption{
		(a) The allowed parameter space of $m_{\phi}$ and $\lambda_{H \phi}$ for $m_{\chi}$ in the range from 5 GeV to 100 GeV. 
		$M_{N} = 100$ GeV  and $y_{DS} = 10^{-12}$. The relic density of $\chi$ is mainly produced by the decay of the freeze-out $\phi$. 
		(b) The allowed parameter space of $m_{\phi}$ and $y_{DS}$ for $m_{\chi}$ in the range from 5 GeV to 100 GeV. 
		$M_{N} = 100$ GeV  and $\lambda_{H \phi} = 0.1$. The relic density of $\chi$ is mainly produced by the freeze-in mechanism. 
	}
	\label{DMrelic2}
\end{figure}

\begin{figure}[!t]
	\centering
	\subfigure[\label{Mphilambdalow}]
	{\includegraphics[width=.486\textwidth]{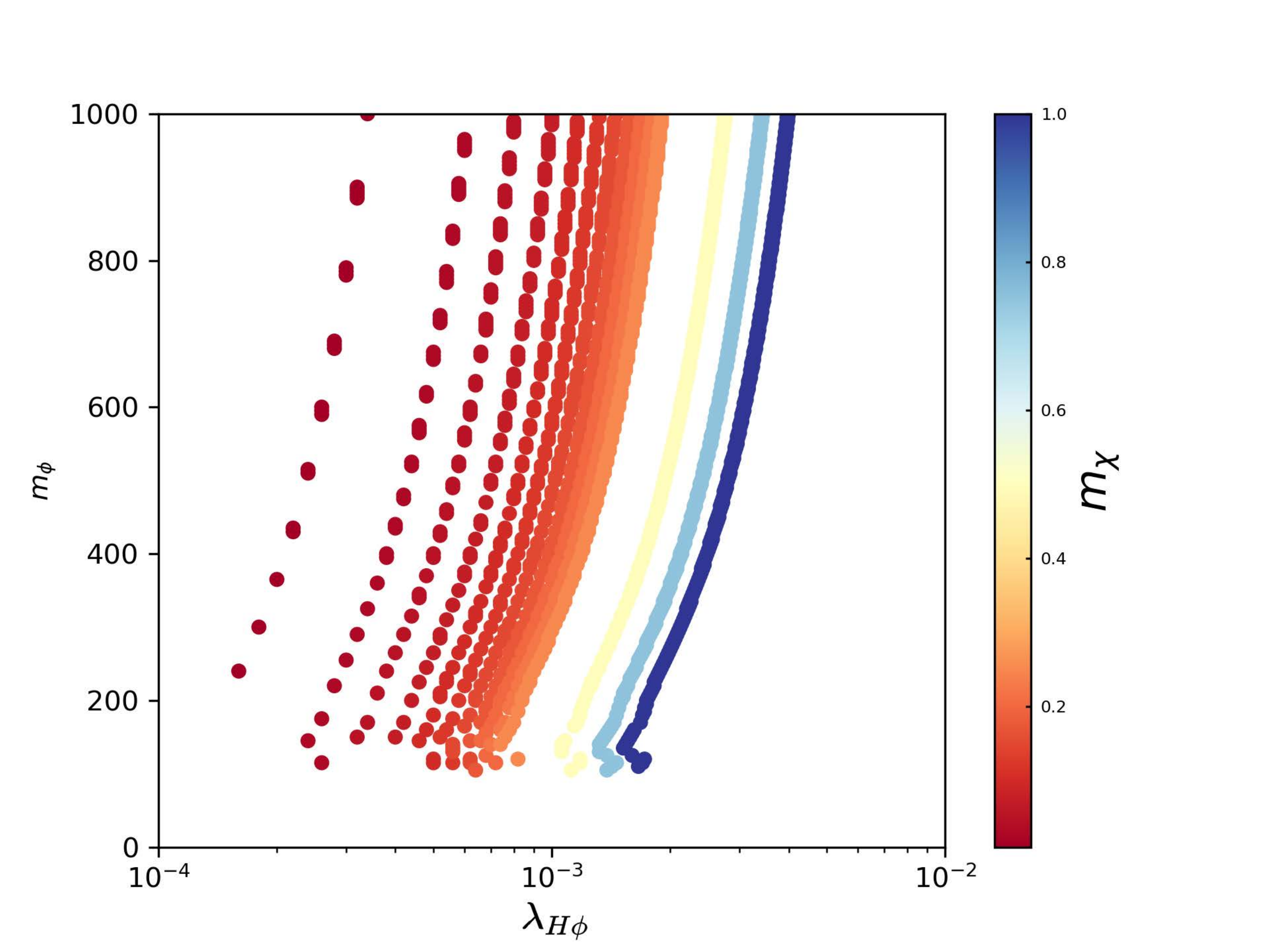}}
	\subfigure[\label{MphiYdslow}]
	{\includegraphics[width=.486\textwidth]{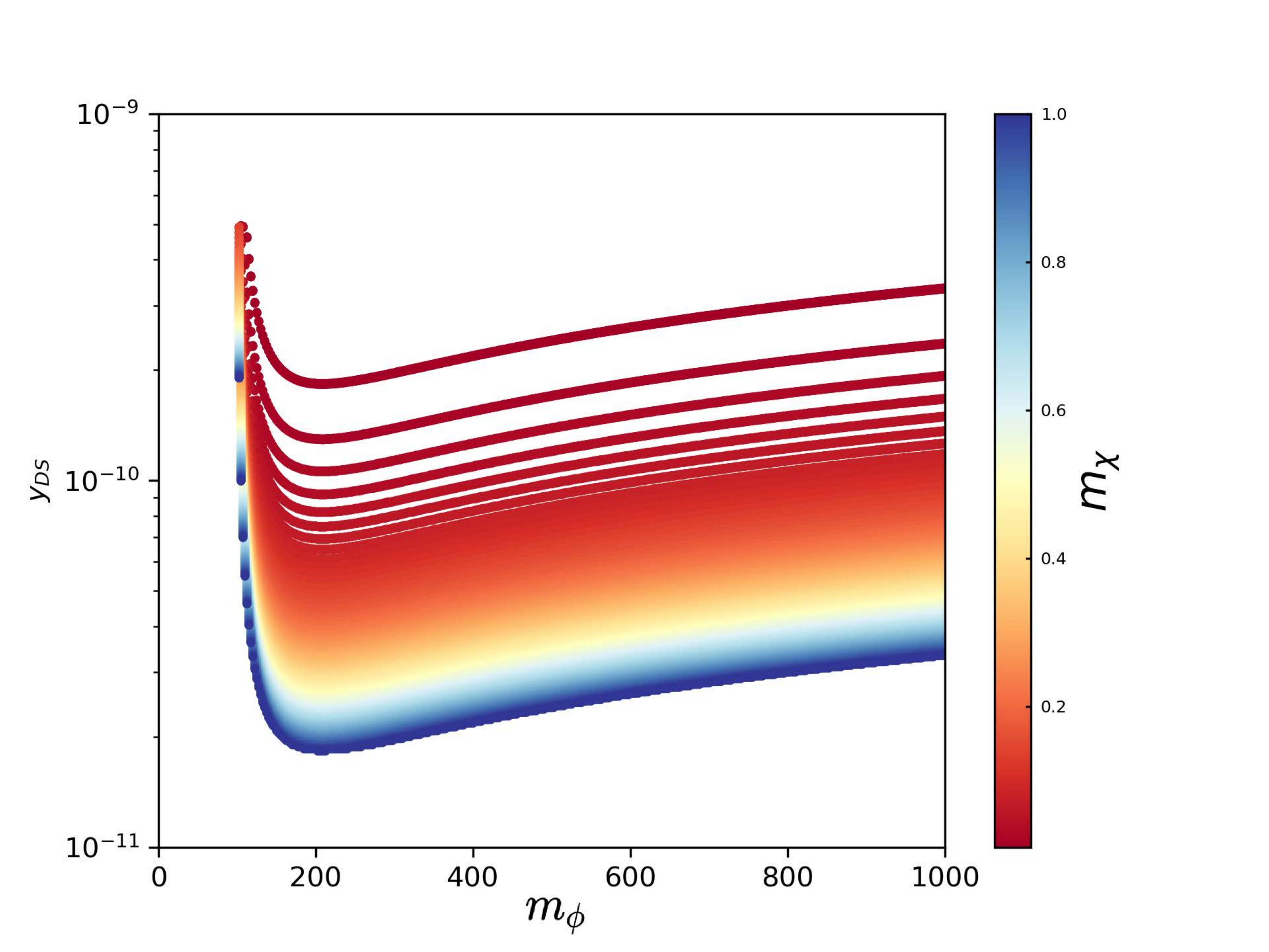}}
	\caption{
	(a) The allowed parameter space of $m_{\phi}$ and $\lambda_{H \phi}$ for $m_{\chi}$ in the range from 10 MeV to 1000 MeV. 
		$M_{N} = 100$ GeV  and $y_{DS} = 10^{-12}$. The relic density of $\chi$ is mainly produced by the decay of the freeze-out $\phi$. 
		(b) The allowed parameter space of $m_{\phi}$ and $y_{DS}$ for $m_{\chi}$ in the range from 10 MeV to 1000 MeV.
		$M_{N} = 100$ GeV  and $\lambda_{H \phi} = 0.1$. The relic density of $\chi$ is mainly produced by the freeze-in mechanism. 
			}
	\label{DMrelic3}
\end{figure}

The allowed parameter space of $\lambda_{H \phi}$ and $y_{DS}$ is shown in Fig.~\ref{yDS_lambda}.
In this figure, we choose $M_{N} = 100$ GeV, $m_{\phi} = 500$ GeV and $m_{\chi} = 10$ GeV. 
We scan the parameter space within the ranges $10^{-13} \leq y_{DS} \leq 10^{-11}$ and $10^{-3} \leq \lambda_{H \phi} \leq 0.1$.
Values of parameters in colored lines are values with which the $\chi$ DM relic abundance is in agreement with the observation. 
The different colors in Fig.~\ref{yDS_lambda} represent different values of $Y_{\phi}/Y_{DM}$ which varies from $0-1$.
which indicates the ratio of the contribution from decay of $\phi$ after freeze out to the total relic density of $\chi$. 

As we can see in the Fig.~\ref{yDS_lambda},
the dominant contribution to the DM relic abundance comes from the decay of $\phi$ after freeze-out
when $\lambda_{H \phi} \sim 0.01$ and $y_{DS} \sim 10^{-12}$. 
On the other hand, the freeze-in mechanism gives a major contribution when $\lambda_{H\phi}$ is larger than about 0.02.
This is because a small $\lambda_{H \phi}$ leads to a large relic density of $\phi$ after freeze-out
which then gives  larger contribution to the $\chi$ DM relic density after $\phi$ decay to $\chi$. 
The opposite situation happens when $\lambda_{H \phi} \sim 0.1$. 
In this case, the $\phi$ relic density is too small to account for the correct $\chi$ DM relic density
and larger freeze-in effect is needed which requires larger $y_{DS}$.
Fig.~\ref{yDS_lambda} also gives a lower bound on $\lambda_{H \phi}$ and 
an upper bound of $y_{DS}$ when $M_{N} = 100$ GeV, $m_{\phi} = 500$ GeV and $m_{\chi} = 10$ GeV.

Fig.~\ref{DMrelic2} and Fig.~\ref{DMrelic3} correspond to two typical scenarios mentioned in Sec.~\ref{sec:freezein}. 
In Fig.~\ref{Mphilambda}, we choose $y_{DS} = 10^{-12}$ GeV and $m_{N} = 100$ GeV.
This corresponds to the scenario shown in Fig.~\ref{sketch1} in which the dominant contribution to the $\chi$ DM relic abundance  
comes from the decay of  the freeze-out $\phi$. 
In the Fig.~\ref{Mphilambda}, we can see that the right relic density can be obtained in a wide range of 
the scalar mass $m_{\phi}$ and the DM particle mass $m_{\chi}$. 
Fig.~\ref{MphiYds} shows the allowed parameter space of $m_{\phi}$ and $y_{DS}$ for the right relic density. 
In Fig.~\ref{MphiYds} we choose $M_{N} = 100$ GeV  and $\lambda_{H \phi} = 0.1$ which corresponds to the situation 
shown in Fig.~\ref{sketch2}. 
We can see that for different $m_{\chi}$ which varies from 10 GeV to 100 GeV, 
$y_{DS}$ can take values from about $2 \times 10^{-12}$ to $10^{-11}$  together  with a right $\chi$ DM relic density.

Fig.~\ref{DMrelic3} shows the allowed parameter space for the right relic density when we take the DM masses from 10 MeV to 1 GeV. Except the DM mass $m_\chi$, other chosen parameters are the same as in Fig.~\ref{DMrelic2}.
 We can see  that  in order to obtain the right $\chi$ DM relic abundance
a smaller value of $\lambda_{H \phi}$ is required in Fig.~\ref{Mphilambdalow} and a bigger value of $y_{DS}$ is required 
in Fig. \ref{MphiYdslow}.   

\section{Conclusions \label{sec:conc}}
In summary, we have studied the possibility that the DM relic density of right magnitude can be produced by the decay of other particles
in the dark sector.  As an example, we have worked in a model which include three right-handed neutrinos, 
one singlet  scalar in the dark sector and one singlet fermion in the dark sector. 
We have studied the possibility that the singlet fermion is lighter and is the DM candidate. 
The dark sector fermion does not couple to the SM directly, but through the right-handed neutrino portal. 
We assume that the coupling of dark sector fermion is very weak so that it never gets into thermal equilibrium with
particles in the SM. The dark sector singlet scalar, on the other hand, can couple to the SM Higgs with sizable interaction
and gets into thermal equilibrium in the early universe. So this singlet scalar would obtain a relic density after its decoupling
through the thermal freeze-out mechanism. Then the decay of this singlet scalar at later time would give rise to a relic density for the singlet fermion
DM. We consider the constraint of DM relic density of right magnitude on the model. 
We find that wide range of  parameter space is allowed with this decay production mechanism.
In particular, the mass of dark matter can be as low as MeV scale while the dark sector singlet scalar can have a
GeV-TeV scale mass and  a weak scale interaction with the SM particles.
So, in this scenario, dark matter  has an interaction much weaker than the
weak scale interaction with the SM,  but the dark sector still has weak scale interaction with the SM.
This is a variant of the WIMP scenario.

We have considered possible constraints on this model of DM.
We assume the singlet scalar to decay at the temperature before reaching MeV scale, in order to avoid possible effect  on BBN.
 From this constraint, we find a lower bound for the coupling of dark sector fermion which is at order $10^{13}-10^{-12}$.
This consideration does not give a strong constraint on the parameter space which gives the right DM relic abundance.
 We have studied the entropy production after the decay of the dark sector singlet scalar , we find that the entropy increase can be neglected.  
 Since the coupling of dark sector fermion is too weak, there are very few restrictions from the direct, indirect DM experiments
 and the collider searches of DM. However, the dark sector singlet scalar still has interaction of weak scale with the SM particles.
 So this model can still be probed in ground based experiments. We leave this topics to future works.
\\
\\

\bigskip
\section*{Acknowledgements}
\label{sec:acknowledgements}
This work is supported by National Natural Science Foundation of China(NSFC), grant No. 11875130.
YC would like to thank Alexander Pukhov for answering the questions on the package \textbf{MicrOmegas}.

\appendix 

\section{Appendix}

\subsection{Cross ections}
\label{App:crosssections}
The cross sections of three major annihilation processes shown in Fig.~\ref{fig:Anndiagrams} take expressions as follows.
\begin{equation}
\label{cs:phiphiNN}
\langle\sigma v\rangle_{\phi \phi \rightarrow N N}=\frac{y_{DS}^{4}m_{N}^{2}}{8 \pi\left(m_{\chi}^{2}+m_{\phi}^{2}-m_{N}^{2}\right)^{2}}\left(1-\frac{m_{N}^{2}}{m_{\phi}^{2}}\right)^{3 / 2}+\mathcal{O}\left(v^{2}\right)
\end{equation}

\begin{equation}
\label{cs:chichiNN}
\langle\sigma v\rangle_{NN \rightarrow \chi \chi   }=\frac{y_{DS}^{4}m_{\chi}^{2}}{32 \pi\left(m_{N}^{2}+m_{\phi}^{2}-m_{\chi}^{2}\right)^{2}}\left(1-\frac{m_{\chi}^{2}}{m_{N}^{2}}\right)^{1 / 2}+\mathcal{O}\left(v^{2}\right)
\end{equation}

\begin{equation}
\langle\sigma v \rangle_{\phi \phi \rightarrow \mathrm{SM}}=\frac{8 \lambda_{H \phi}^{2} v_{ew}^{2}}{\left(4 m_{\phi}^{2}-m_{h}^{2}\right)^{2}+m_{h}^{2} \Gamma_{h}^{2}} \frac{\Gamma_{h \rightarrow \mathrm{SM}}\left(m_{h}=2 m_{\phi}\right)}{2 m_{\phi}}
\end{equation}
where $v_{E W}=246$ $\mathrm{GeV}$ and $\Gamma_{h \rightarrow \mathrm{SM}}\left(m_{h}=2 m_{\phi}\right)$ denotes the
width of the Higgs boson decays into SM particles for a Higgs mass of $2 m_{\phi}$.

\subsection{Heavy neutrino decay widths}\label{App:Ndecaywidths}
 $m_N$  larger than $m_W$, $m_Z$ or $m_h$, 
the corresponding two body decay widths can be be written as\cite{Atre:2009rg,He:2009mv,Banerjee:2015gca,delAguila:2008cj,Pilaftsis:1991ug}:
\begin{equation}
\Gamma\left(N \rightarrow W^{\pm} \ell^{\mp}\right)=\frac{g^{2}}{64 \pi}\left|R_{l N}\right|^{2} \frac{m_{N}^{3}}{m_{W}^{2}}\left(1-\frac{m_{W}^{2}}{m_{N}^{2}}\right)^{2}\left(1+\frac{2 m_{W}^{2}}{m_{N}^{2}}\right)
\end{equation}
\begin{equation}
\Gamma\left(N \rightarrow Z v_{l}\right)=\frac{g^{2}}{64 \pi c_{W}^{2}}\left|R_{l N}\right|^{2} \frac{m_{N}^{3}}{m_{Z}^{2}}\left(1-\frac{m_{Z}^{2}}{m_{N}^{2}}\right)^{2}\left(1+\frac{2 m_{Z}^{2}}{m_{N}^{2}}\right)
\end{equation} 
\begin{equation}
\Gamma\left(N \rightarrow h v_{l}\right)=\frac{g^{2}}{64 \pi}\left|R_{l N}\right|^{2} \frac{m_{N}^{3}}{m_{W}^{2}}\left(1-\frac{m_{h}^{2}}{m_{N}^{2}}\right)^{2}
\end{equation}
Note that $R_{l N_{j}}$ is the mixing matrix element of the active neutrinos in the flavor base $\nu_{l}(l=e, \mu, \tau)$ with heavy neutrinos $N_j$.
 $\nu_{l}$ can be expressed by a mixture of
the light neutrinos in mass eigenstates $\nu_{i}$ and
heavy sterile neutrinos in mass eigenstates $N_j$:
\begin{equation}
\nu_{l}=\sum_{i} U_{l i} \nu_{i}+\sum_{j} R_{l N_{j}} N_{j},
\end{equation}
where $U_{lj}$ is the Pontecorvo-Maki-Nakagawa-Sakata (PMNS) mixing matrix. 
 For small enough $\left|R_{l N_{j}}\right|$, the PMNS mixing matrix $U$ can be considered as approximately unitary.

The tree level decay rate of sterile neutrino decaying to three final fermions through interaction with Z and W bosons induced by mixing with active neutrinos are summarized here\cite{Liao:2017jiz}. 
Effects of on-shell and off-shell Z and W bosons are all taken into account by including the width of W and Z in the propagators.
This formalism can be used for general parameter space of $m_N$, not just for $m_N < m_W$\cite{Liao:2017jiz}.
\\(1) For $N \to l_1^- l_2^+ \nu_{l_2}$, $N \to l_1^+ l_2^- {\bar \nu}_{l_2}$ and $l_1\neq l_2$
\begin{eqnarray}
\Gamma(N \to l_1^- l_2^+ \nu_{l_2})=\Gamma(N \to l_1^+ l_2^- {\bar \nu}_{l_2})
=|R_{l_1 N}|^2\frac{G_F^2 m_N^5}{\pi^3} F_N(m_N,m_W,\Gamma_W), \label{decayR1}
\end{eqnarray}
where $F_N$ is a dimensionless function and is given in (\ref{FN}) below. 
\\(2) For $N \to l^- q_1 {\bar q}_2$, $N \to l^+ {\bar q}_1 q_2$
\begin{eqnarray}
\Gamma(N\to  l^- q_1 {\bar q}_2)=\Gamma(N\to l^+ {\bar q}_1 q_2)
=|R_{l N}|^2\frac{G_F^2 m_N^5}{\pi^3} N_C F_N(m_N,m_W,\Gamma_W) |K_{q_1 q_2}|^2. \label{decayR2}
\end{eqnarray}
$K_{q_1 q_2}$ is the CKM matrix element in $(q_1,q_2)$ entry, $N_C=3$ the number of color degrees of freedom of quarks. 
\\(3) For $N\to l^- l^+ \nu_l$, $N\to l^+l^- {\bar \nu}_l$
\begin{eqnarray}
&&\Gamma(N\to l^- l^+ \nu_l)=\Gamma(N\to l^+ l^- {\bar \nu}_l) \nonumber \\
&& =|R_{l N}|^2\frac{G_F^2 m_N^5}{\pi^3} [ F_N(m_N,m_W,\Gamma_W)+(C_L^2+C_R^2)F_N(m_N,m_Z,\Gamma_Z) \nonumber \\
&&+2 C_L  ~F_S(m_N, m_W,\Gamma_W,m_Z,\Gamma_Z)], \label{decayR3}
\end{eqnarray}
where $C_{L,R}$ is given in (\ref{couplings0-1}), $F_S$ is a dimensionless function and is given below in (\ref{FS}). 
\\(4) For $N\to \nu_l {\bar l}' l'$ and $N\to {\bar \nu}_l l' {\bar l}'$
\begin{eqnarray}
\Gamma(N\to \nu_l {\bar l}' l')=\Gamma(N\to {\bar \nu}_l l' {\bar l}')=
|R_{l N}|^2\frac{G_F^2 m_N^5}{\pi^3} (C_L^2+C_R^2) F_N(m_N,m_Z,\Gamma_Z). \label{decayR4}
\end{eqnarray}
\\(5) For $N \to \nu_l q {\bar q}$ and $N \to {\bar \nu}_l {\bar q} q$
\begin{eqnarray}
\Gamma(N \to \nu_l q {\bar q})=\Gamma(N \to {\bar \nu}_l {\bar q} q)=
|R_{l N}|^2\frac{G_F^2 m_N^5}{\pi^3}N_C [(C^q_L)^2+(C^q_R)^2] F_N(m_N,m_Z,\Gamma_Z), \label{decayR5}
\end{eqnarray}
where $q=u,d,c,s,b$ for $m_N < 2 m_t$ and $C^q_{L,R}$ is given in (\ref{couplings0-2}) and (\ref{couplings0-3}).
\\(6) For $N \to \nu_l {\nu}_{l'} {\bar \nu}_{l'}$ and $N \to {\bar \nu}_l {\bar \nu}_{l'}  {\nu}_{l'} $, $l\neq l'$
\begin{eqnarray}
\Gamma(N \to \nu_l {\nu}_{l'} {\bar \nu}_{l'})=\Gamma(N \to {\bar \nu}_l {\nu}_{l'} {\bar \nu}_{l'})
=|R_{l N}|^2\frac{G_F^2 m_N^5}{\pi^3} C_\nu^2 F_N(m_N,m_Z,\Gamma_Z), \label{decayR6}
\end{eqnarray}
where $C_\nu =1/2$.
\\(7) For $N \to \nu_l {\nu}_l {\bar \nu}_l$ and $N \to {\bar \nu}_l {\bar \nu}_l {\nu}_l $
\begin{eqnarray}
\Gamma(N \to \nu_l {\nu}_{l} {\bar \nu}_{l})=\Gamma(N \to {\bar \nu}_l {\nu}_{l} {\bar \nu}_{l})
=|R_{l N}|^2\frac{G_F^2 m_N^5}{\pi^3} 4 C_\nu^2 F_N(m_N,m_Z,\Gamma_Z). \label{decayR7}
\end{eqnarray}

Couplings $C_L$, $C_R$ etc. which appear in expressions above, are given as
\begin{eqnarray}
& C_L=-\frac{1}{2}+\sin^2\theta_W,~C_R=\sin^2\theta_W,\label{couplings0-1} \\
& C_L^u=\frac{1}{2}-\frac{2}{3} \sin^2\theta_W, C_R^u=-\frac{2}{3} \sin^2\theta_W, \label{couplings0-2} \\
& C_L^d=-\frac{1}{2}+\frac{1}{3} \sin^2\theta_W, C_R^d=\frac{1}{3} \sin^2\theta_W. \label{couplings0-3}
\end{eqnarray}
For mass $m_N$, $m_X$ and decay rate $\Gamma_X$, the function $F_N$ used above is
\begin{eqnarray}
F_N=\frac{1}{m_N^4}\int^{m_N \over 2}_0 dE_1 \int^{m_N \over 2}_{{m_N \over 2}-E_1}dE_2|X_W|^2 
\frac{1}{2}(m_N-2 E_2) E_2, \label{FN}
\end{eqnarray}
where $X_W$ comes from the propagator of W boson and is
\begin{eqnarray}
X_W=\frac{m_W^2}{q^2-m_W^2+i\Gamma_W m_W},\label{XW}
\end{eqnarray}
where $q^2=m^2_N-2 m_N E_1$ and $\Gamma_W$ is the total decay rate of $W$. 
As shown in \cite{Liao:2017jiz}, $F_N$ can be obtained as an analytic form.

Function $F_S$ in (\ref{decayR3}) is given as
\begin{eqnarray}
F_S=\frac{1}{m_N^4}\int^{m_N \over 2}_0 dE_1 \int^{m_N \over 2}_{{m_N \over 2}-E_1}dE_2(X_W X_Z^* +X_W^* X_Z) 
\frac{1}{2}(m_N-2 E_2) E_2, \label{FS}
\end{eqnarray}
where
\begin{eqnarray}
X_Z=\frac{m_Z^2}{q^2_3-m_Z^2+i\Gamma_Z m_Z}.\label{XZ}
\end{eqnarray}
$q_3^2=m_N^2-2 m_N E_3$ with $E_3=m_N-E_1-E_2$ when considering
the decay of $N$ at rest. In these results, masses of the final fermions have all been neglected.

For $m_N$ greater than $m_H$, the mass of Higgs boson, $N$ can also decay to $H$ via $N \to \nu({\bar \nu})H$. The partial decay widths here can be similarly obtained with the introduction of a function
$F_N(m_N,m_H,\Gamma_H)$.  For example, for $N\to \nu_l {\bar f} f$ and $N \to {\bar \nu}_l {\bar f} f$
\begin{eqnarray}
\Gamma(N\to  \nu_l {\bar f} f)=\Gamma(N \to {\bar \nu}_l {\bar f} f)
=\frac{g^2 m_N^7 |R_{lN}|^2 y_f^2}{16\pi^3 m_W^2 m_H^4} N_f F_N(m_N,m_H,\Gamma_H) , \label{decayR-NtoH}
\end{eqnarray}
where $y_f$ is the Yukawa coupling of fermion $f$, $N_f=1$ for f  being a lepton and $N_f=3$ for f being a quark.

\bibliography{darkmatter}

\bibliographystyle{utphys}

\providecommand{\href}[2]{#2}\begingroup\raggedright\endgroup

\end{document}